\documentclass[a4paper,twocolumn]{article}
\usepackage[DIV=15,BCOR=2mm,headinclude=true,footinclude=false]{typearea}

\usepackage{subcaption}
\usepackage{amssymb}

\usepackage{mathptmx}
\usepackage{amsmath}
\usepackage{caption}
\usepackage{subcaption}
\usepackage{graphicx}
\usepackage{lineno}
\usepackage{xurl}

\usepackage[breaklinks=true,linktocpage,breaklinks]{hyperref}
\usepackage{breakcites}

\usepackage{authblk}
\author[1,2,*]{Karsten Schuhmann}
\author[1,2]{Klaus Kirch}
\author[2]{Miroslaw Marszalek}
\author[3]{Francois Nez}
\author[4]{Randolf Pohl}
\author[1]{Ivo Schulthess}
\author[2]{Laura Sinkunaite}
\author[1]{Gunther Wichmann}
\author[1]{Manuel Zeyen}
\author[1,2]{Aldo Antognini}

\affil[1]{Institute for Particle Physics and Astrophysics, ETH,  8093 Zurich, Switzerland}
\affil[2]{Paul Scherrer Institute, 5232 Villigen PSI, Switzerland}
\affil[3]{Laboratoire Kastler Brossel, UPMC-Sorbonne Universit\'es, CNRS,
ENS-PSL Research University, Coll\`ege de France, 75005 Paris, France}
\affil[4]{Johannes Gutenberg-Universit\"at Mainz, QUANTUM, Institut f\"ur Physik \& Exzellenzcluster PRISMA, 55128 Mainz, Germany}
\affil[*]{Corresponding author: skarsten@phys.ethz.ch}

\title{Multi-pass amplifiers with self-compensation of the thermal lens}

\begin{document}

\maketitle
\begin{abstract}
We present a novel architecture for a multi-pass amplifier  based on a succession of optical Fourier transforms and short propagations that shows a superior stability for variations of the thermal lens compared to state-of-the-art 4f-based amplifiers.
We found that the proposed multi-pass amplifier is robust to variations of the active medium dioptric power.
The superiority of the proposed architecture is demonstrated by analyzing the variations of the size and divergence of the output beam in form of a Taylor expansion around the design value for variations of the thermal lens in the active medium.
%
%
The dependence of the output beam divergence and size is investigated also for variations of the number of passes, for aperture effects in the active medium and as a function of the size of the beam on the active medium. 
This architecture makes efficient use of the transverse beam filtering inherent in the active medium to deliver a beam with excellent quality (TEM00).
\end{abstract}

\section{Introduction}

Multi-pass amplifiers are used for the generation of the currently highest laser pulse energies and average powers~\cite{Keppler2013,Koerner2011,Korner2016, Jung2016, Siebold2016, Negel:15}.
Oscillators, regenerative amplifiers and multi-pass amplifiers are limited in pulse energy by the damage of optical components induced by the high intensity.
Multi pass amplifiers possess a higher damage threshold compared to laser oscillators and regenerative amplifiers due to the absence of intra cavity enhancement and non-linear crystals~\cite{Butze:04,  Martinez1998}.
Optical damage can be avoided by increasing the laser beam size, but this typically increases the sensitivity of the beam propagation and output beam characteristics (size and divergence) to the active medium thermal lens.
The active medium design  and material choice  can be optimized to decrease the thermal lens, e.g., in thin-disk lasers~\cite{Giesen1994, Brauch:95, Stewen2000, Peters2009}. 
Still, the higher pump powers and the increased beam size needed for energy and power scaling, make the thermal lens in the active medium the most severe limitation to be overcome~\cite{Chvykov:16,  Negel:15,Vaupel:13}.
For these reasons an optical layout has to be chosen that minimizes the changes of the beam propagation for  variations of the thermal lens of the active medium.

Commonly, relay imaging (4f-imaging) from active medium to active medium is used to realize a multi-pass amplifier where the output power is independent of the thermal lens of the active medium because the beam is imaged from pass to pass~\cite{Hunt:78}.
The 4f-design exhibits various advantages: It sustains equal beam size at each pass on the active medium independently of the dioptric power of the active medium and independently of the beam size at the active medium. 
Moreover, numerous passes can be realized with only a few optical elements~\cite{Georges:91, Wojtkiewicz:04, Kaksis:16, Plaessmann:93, Lundquist2010, Plaessmann1992, Erhard2000,Scott:01, Papadopoulos:11, Banerjee:12, Neuhaus:08}.
Yet, this design does not represent the optimal solution for a multi-pass amplifier because the phase front curvature of the beam after propagation through this optical system depends strongly on variations of the active medium dioptric power.

In this paper we present a novel multi-pass architecture whose propagation from pass to pass is given by a succession of Fourier transforms and short propagations. 
The working principle, the practical implementation and the advantages in term of stability to variations of dioptric power (thermal lens) and aperture effects of the active medium are detailed.

\section{Thermal lens and aperture effects}
\label{Thermal lens and aperture effects}

In this section, we briefly review well-known facts and various definitions 
needed
in the following sections.

When a laser beam crosses the pumped active medium, it is amplified and it experiences a position-dependent optical phase delay (OPD) that distorts its phase front curvature.
When simulating the propagation of the laser beam, this distortion can be mostly accounted for by approximating the active medium with a spherical lens~\cite{Osterink68, Koechner:70} that can be easily described within the ABCD-matrix formalism~\cite{Kogelnik:66}.

The position-dependent gain (and absorption in the unpumped region) can be approximated by an average gain per pass and a position-dependent transmission function that shows a gradual transition from maximal transmission (including gain) on the optical axis to zero transmission far from the axis.
%
This position dependent gain is commonly referred to as "soft aperture".
 In this study we consider only Gaussian beams in the fundamental mode and we approximate the super-Gaussian aperture of the pumped active medium as Gaussian aperture.
 This approximation is justified because the Fourier transform (that underlies the here presented designs) converts the high-frequency spatial distortions of the beam caused by the super-Gaussian aperture into a halo of the fundamental mode that is eliminated in the next pass at the active medium. 
Therefore, the beam propagating in the amplifier shows only minor higher-order distortions from the fundamental mode so that the description of the super-Gaussian aperture can be effectively accomplished using a Gaussian aperture~\cite{PHDSchuhmann}. 
In fact, a Gaussian aperture transforms a Gaussian input beam into a Gaussian output beam~\cite{siegman1986lasers}.
The relation between the $1/e^2$-radii of the input ($w_\mathrm{in}$) and output ($w_\mathrm{out}$) beams takes the simple form
\begin{equation}
  w^{-2}_\mathrm{out}=w^{-2}_\mathrm{in}+W^{-2} ,
  \label{eq:aperture-decrease}
\end{equation}
where $W^2$ is the $1/e^2$-radius of the intensity transmission function of the Gaussian aperture~\cite{siegman1986lasers,Eggleston:81}.
This simple behavior of a Gaussian aperture acting on a Gaussian beam (TEM00-mode) can be captured into an ABCD-matrix for Gaussian optics as~\cite{Kogelnik:65,siegman1986lasers,Bowers:92,Andrews:95,Kalashnikov:97,PHDSchuhmann}:
\begin{eqnarray}
  M_\mathrm{aperture}=\left[
    \begin{array}{cc} 1 & 0 \\ 
      -i\frac{\lambda}{\pi W^2} & 1
    \end{array}
\right]\ ,
\end{eqnarray}
where $\lambda$ is the  wavelength of the laser beam.

An active medium (AM) usually exhibits both  a lens effect (position-dependent OPD) and an aperture effect (position-dependent gain).
These can be merged into a single ABCD-matrix describing the active medium $M_\mathrm{AM}=M_\mathrm{lens} \cdot M_\mathrm{aperture}$, where $M_\mathrm{lens}$ is the ABCD-matrix of a thin spherical lens.
The product of these two ABCD-matrices becomes
\begin{eqnarray}
  M_\mathrm{AM}=
  \left[
    \begin{array}{cc} 1 & 0 \\ 
      \hspace{-2mm}-V & 1
    \end{array}
    \right] \hspace{-1mm}\cdot\hspace{-1mm}
  \left[
    \begin{array}{cc} 1 & 0 \\ 
      \hspace{-2mm}-i\frac{\lambda}{\pi W^2} & 1
    \end{array}
\right]= \left[
    \begin{array}{cc} 1 & 0 \\ 
      \hspace{-2mm}-V-i\frac{\lambda}{\pi W^2} & 1
    \end{array}
    \right],
  \label{eq:matrix-2}
\end{eqnarray}
where $V$ is the dioptric power of the active medium.
In Eq.~(\ref{eq:matrix-2}) we have assumed that the length of the active medium is negligible, i.e., that the length of the active medium is much smaller than the Rayleigh length of the beam.
This assumption is generally well justified for high-power lasers given the large beam waists they have at the active medium.

The similarity of $M_\mathrm{AM}$ and $M_\mathrm{lens}$ allows us to define a complex dioptric power $\tilde{V}$ ($\tilde{V} \in \mathbb{C}$) of the active medium as
\begin{equation}
  \tilde{V} =V+i\frac{\lambda }{\pi W^2}.
\end{equation}
Its real part $V$ is the ``standard'' dioptric power (focusing-defocusing effects), while the imaginary part accounts for the Gaussian aperture.
The real part can be articulated as a sum of various contributions
\begin{equation}
    V = V_\mathrm{unpumped} + V_\mathrm{thermal}
      = V_\mathrm{unpumped} + V_\mathrm{thermal}^\mathrm{avg} + \Delta V \ ,
\end{equation}
where $V_\mathrm{unpumped}$ is the dioptric power of the unpumped active medium and $V_\mathrm{thermal}$ the dioptric power related to the pumping of the active medium (thermal-lens effect).
The latter can be expressed by a known average a value $V_\mathrm{thermal}^\mathrm{avg}$ and an unknown deviation $\Delta V$ from this average value that might depend among others on running conditions and imperfection of the active medium due to the manufacture process.

In Secs.~\ref{sec:state-of-the-art}-\ref{sec:many-passes}, for simplicity, but without loss of generality, we assume that $V_\mathrm{unpumped}+V_\mathrm{thermal}^\mathrm{avg}=0$.
In Sec.~\ref{sec:implementation}, when realistic implementations of multi-pass amplifiers are presented, we relax this assumption.
Moreover, in the various plots presented in this paper we have fixed the value of the aperture size to be $W = 4w_\mathrm{in}$.
For comparison, in some figures we also present results for $W = \infty$, to highlight the aperture effects.
 
A laser mode size of $w_\mathrm{in} \approx 0.7R_p$, where $R_p$ is the radius of the super-Gaussian pump spot, is typically used in the thin-disk laser community for efficient laser operation in the fundamental mode.
The above choice of a Gaussian aperture with $W=4w_\mathrm{in}$ rest upon the fact that it produces a similar reduction of the fundamental mode size and a similar reduction of the fundamental mode transmission compared to a super-Gaussian aperture fulfilling the relation $w_\mathrm{in} \approx 0.7R_p$ ~\cite{PHDSchuhmann}.

\section{State-of-the-art multi-pass amplifiers}
\label{sec:state-of-the-art}

\begin{figure}[t!]
\centering
  \begin{subfigure}[t]{.9\linewidth}
    \centering
    \includegraphics[width=\linewidth]{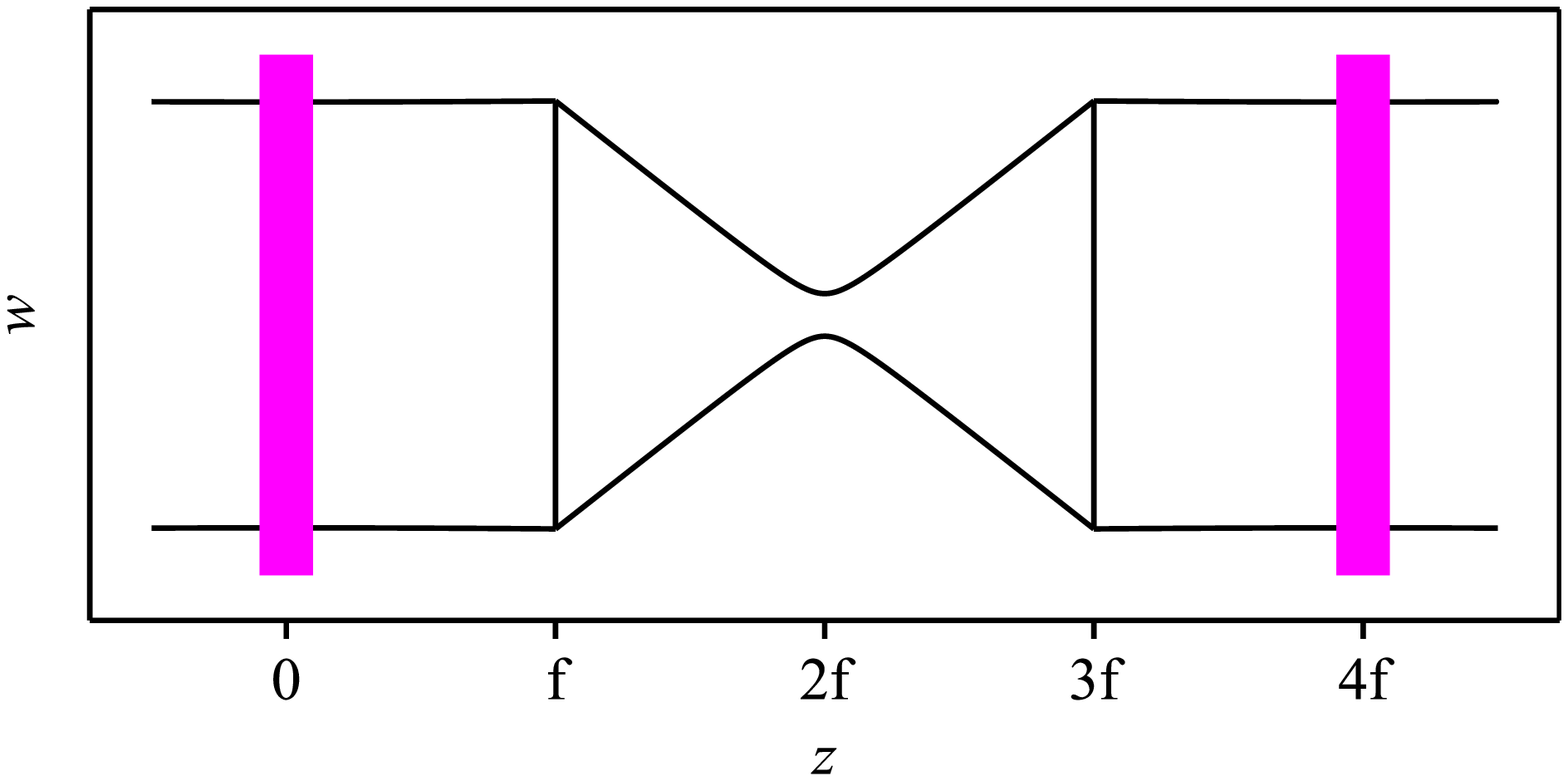} 
    \caption{} \label{fig:4f-segment-a}
  \end{subfigure}
  \begin{subfigure}[t]{.9\linewidth}
    \centering
    \includegraphics[width=\linewidth]{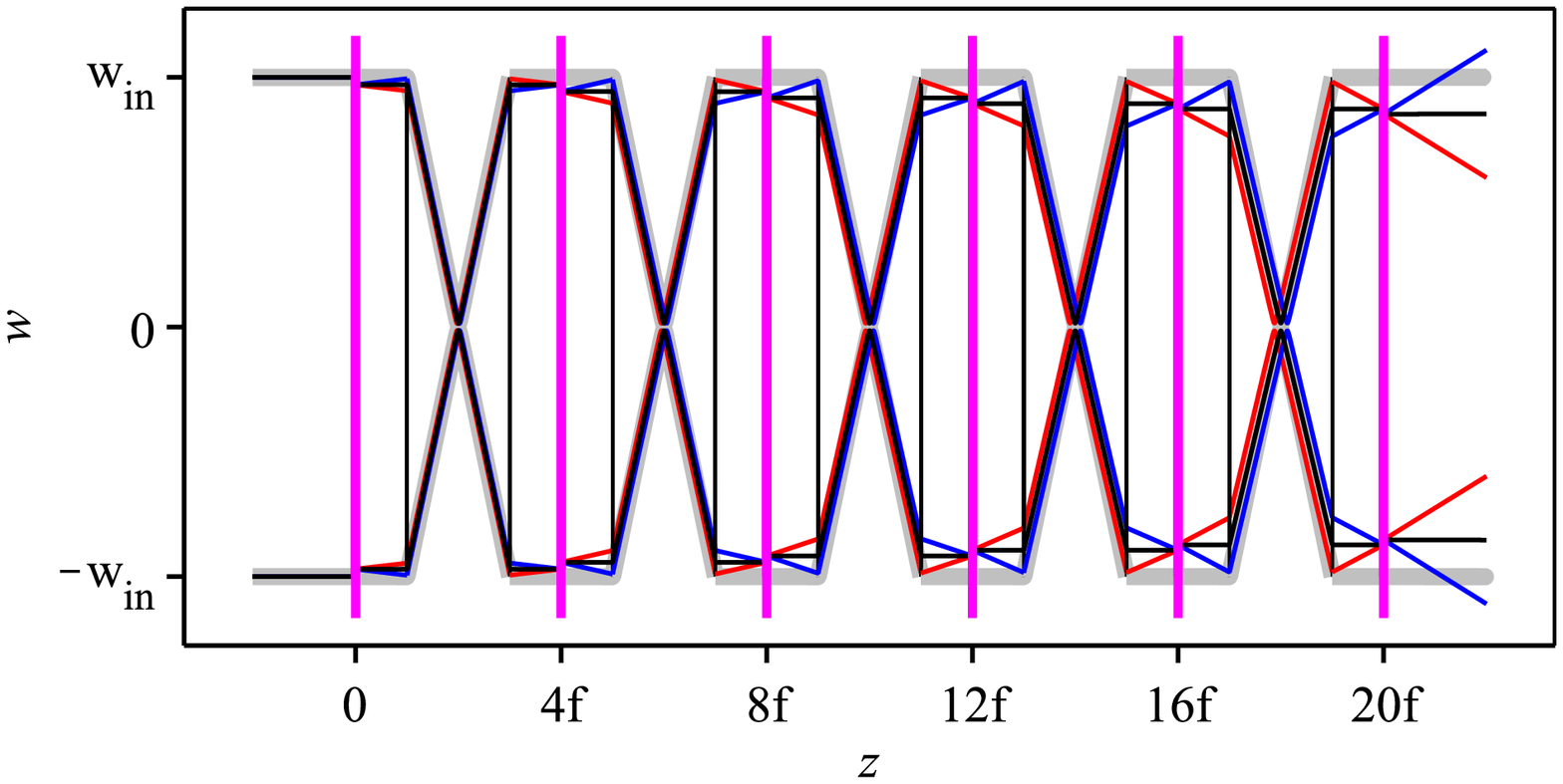} 
    \caption{} \label{fig:4f-segment-b}
  \end{subfigure}
\caption{(a) Scheme of the laser beam propagation from active medium to active medium through a 4f-imaging segment.  
The 4f-segment is  given by a free propagation of distance f, a focusing optics with  focal length f, a free propagation with length 2f, a second  focusing element with focal length f, and a free propagation of length f. 
The vertical magenta lines represent the position of the active medium; the vertical black lines indicate the  position of the optic with focal length f. 
The black curves  represent the beam size evolution along the optical axis $z$. 
(b)  6-pass amplifier and corresponding beam size evolution along the  propagation axis realized by concatenating five 4f-imaging stages  and assuming a Gaussian aperture at the active medium of  $W=4w_\mathrm{in}$.  
The black curves show the beam size evolution for an  active medium dioptric power having the design value, i.e. $V  = 0$. 
The red and blue curves show the propagations (beam size evolution) for   $V=\pm\frac{1}{40 f}$. The gray curves represent the beam evolution for $V=0$ and $W=\infty$.
The discontinuity of the waist size at the gain medium is caused by the (Gaussian) aperture effect of the finite pump region.}
\label{fig:4f-segment}
\end{figure}

The commonly used state-of-the-art multi-pass amplifiers are based on 4f relay imaging (4f) from active medium (AM) to active medium~\cite{Georges:91, Wojtkiewicz:04, Kaksis:16, Plaessmann:93, Lundquist2010, Plaessmann1992, Erhard2000,Papadopoulos:11, Banerjee:12,Scott:01, Hunt:78}, so that in these amplifiers the beam proceeds as a succession of propagations
\begin{equation*}
\noindent\hspace*{5mm} \cdots AM-4f-AM-4f-AM-4f-AM-4f-AM\cdots .
\end{equation*}
The geometrical layout and the corresponding propagation of a laser beam through a 4f segment is given in Fig.~\ref{fig:4f-segment-a}.
By concatenating for example six of such propagations, the six-pass amplifier of Fig.~\ref{fig:4f-segment-b} can be realized.

It can be demonstrated that the 4f-imaging warrants the relation $w_\mathrm{in, \, N}=w_\mathrm{out, \, N-1}$~\cite{Hunt:78}.
Here, we defined $w_\mathrm{in,\, N}$ as the size of the beam incident on the $N$-th pass at the active medium, and $w_\mathrm{out,\, N}$ as the size of the beam departing from the $N$-th pass at the active medium after amplification.
In the absence of aperture effects ($W=\infty$) in the active medium $w_\mathrm{out, \, N}=w_\mathrm{in, \, N}$. 
Hence, in the absence of aperture effects the beam size is identically reproduced at each pass on the active medium independently of the size of the laser beam.
This property, together with the simplicity of the implementation (many passes using few optical elements), are the main advantages of the 4f-based multi-pass architecture.
However, for non-vanishing aperture effects ($W \neq \infty$) 
the beam size is decreased when passing the active medium $w_\mathrm{out, \, N}<w_\mathrm{in,  \, N}$ (see Eq.~(\ref{eq:aperture-decrease})).  
Because the 4f-imaging reproduces the beam $w_\mathrm{in, \, N+1}=w_\mathrm{out,  \, N}$ the beam size size is continuously decreased at each pass $w_\mathrm{out, \, N}<w_\mathrm{out,  \, N+1}$ as visible in Fig.~\ref{fig:4f-segment-b}.

By considering the various curves presented in Fig.~\ref{fig:4f-segment-b} another, even more crucial, drawback of the 4f-based architecture emerges: namely that the divergence of the output beam strongly depends on variations of the dioptric power $\Delta V$ of the active medium.

The output beam characteristics and their dependence on variations of  $\Delta V$, can be derived by making use of the ABCD-matrix and  complex beam parameter formalism~\cite{Kogelnik:65}.
The complex beam parameter $q$ is defined as
\begin{equation}
  \frac{1}{q}=\frac{1}{R}-i \frac{\lambda}{\pi w^2} \ ,
  \label{eq:complex-beam-parameter}
\end{equation}
where $R$ is the radius of curvature of the phase front and $w$ the $1/e^2$-radius of the TEM00 Gaussian beam.
In this framework the complex beam parameter at the output of an optical system ($q_\mathrm{out}$), can be computed knowing the ABCD-matrix of the optical system and the complex beam parameter at its input ($q_\mathrm{in}$) using  the relation~\cite{Kogelnik:65}
\begin{equation}
  q_\mathrm{out}=\frac{Aq_\mathrm{in}+B}{Cq_\mathrm{in}+D} \ . 
  \label{eq:complex-beam-evolution}
\end{equation}

The ABCD-matrix describing the Gaussian beam propagation from the first pass to the last pass of a $N$-pass amplifier having ($N-1$) 4f-imaging propagations reads
\begin{equation}
M_\mathrm{4f,\, N}=\left[ \begin{array}{l l}
    {\left(-1\right)}^{N-1} & 0 \\ {\left(-1\right)}^N\cdot N\cdot
    \Delta \tilde{V} & {\left(-1\right)}^{N-1} \end{array} \right] \ .
\label{eq:4f-ABCD}
\end{equation}
The complex beam parameter of the output beam just after the last pass can be obtained by applying Eq.~(\ref{eq:complex-beam-evolution}) to the ABCD-matrix of Eq.~(\ref{eq:4f-ABCD}).
Its Taylor expansion around $q_\mathrm{in}$ for small variations of $\Delta \tilde{V}$ takes the form
\begin{equation}
  q_\mathrm{out, \, N} = q_\mathrm{in} + N q^2_\mathrm{in} \Delta \tilde{V} + N^2
  q_\mathrm{in}^3{ \Delta \tilde{V}}^2 + N^3 q_\mathrm{in}^4 {\Delta \tilde{V}}^3 + \cdots
  \  .
  \label{eq:Taylor-4f}
\end{equation}

Notably, this expansion shows a linear dependence of $q_\mathrm{out, \, N}$ to variations of $\Delta \tilde{V}$, implying a sensitivity of the output beam characteristics also to small variations of $\Delta \tilde{V}$.
In addition, the linear dependence in $N$ indicates that there is an accumulation of aperture and lens effects at each pass though the active medium (see the increase of beam divergence and decrease of beam size at each pass in Fig.~\ref{fig:4f-segment-b}).

From Eq.~(\ref{eq:Taylor-4f}) it is possible to extract the phase front curvature and size of the output beam.
When aperture effects are neglected ($W = \infty$)  they are given by
\begin{eqnarray}
  w_\mathrm{out, \, N}                &= &w_\mathrm{in}\\
  R^{-1}_\mathrm{out, \, N} & = & N \,\Delta V \ .
  \label{eq:4f-R} 
\end{eqnarray}
While the output beam size does not depend on variations of the dioptric power $\Delta V$, the phase front curvature of the output beam shows the above mentioned accumulation of the dioptric power deviation $\Delta V$, leading to an increasing deviation from the design value of $R^{-1}_\mathrm{out, \, N}=0$ with increasing number of passes.
This implies a strong dependence of the output beam divergence to variations of the operational conditions.

When aperture effects are included, the beam size right after the $N$-th pass on the active medium shrinks with increasing $N$ as:
\begin{equation}
  \frac{w_\mathrm{out, \, N}}{w_\mathrm{in}} =  1
    -\frac{w_\mathrm{in}^{2}  N}{2\,W^2}
    +\frac{3\,w_\mathrm{in}^4  N^2}{8\,W^4}
    -\frac{5\, w_\mathrm{in}^6 N^3}{16\, W^6}+ \cdots \; .
  \label{eq:4f-R-b} 
\end{equation}
Here, for simplicity and to better discern the dependence of $w_\mathrm{out, \, N}$ on the aperture size $W$ we have assumed that $V=\Delta V=0$.

Given the imaging properties of the 4f-scheme, also distortions of the $OPD$ beyond what is accounted for by the spherical lens, as well as distortions caused by the deviation from a purely Gaussian aperture, are accumulated at each pass.
Thus, while propagating in the 4f-based multi-pass amplifier the beam shows increasing mode deformation (beyond the TEM00-mode) resulting in an output beam with decreased beam quality and increased dependency on running conditions~\cite{Keppler2012, Pittman2002}.

For 4f-based amplifiers there are two methods to reduce the accumulation of phase front distortions.
The first is by changing the distance between the two lenses used for the imaging (see Fig. 1a)~\cite{Kaksis:16, friebel:hal-01222067}, so that their separation departs from 2f. 
To put it in another way, the distance between the focusing elements of the 4f-propagation can be adjusted to preserve the $q$-parameter also for a curved active medium~\cite{Neuhaus2009Passi-9533}.
The second approach is by realizing the N-pass 4f-amplifier as a succession of 2N 4f-segments so that between each pass on the active medium there is an intermediate image of the active medium.
At the position of this intermediate image, a deformable mirror can be placed adjusted to produce an opposite phase front distortion compared to the active medium \cite{Keppler2013}.
These methods are particularly suited  to compensate for non-zero values of the sum $V_\mathrm{unpumped} + V_\mathrm{avg}$ .
For time-dependent variations of  $\Delta V$, a time dependent correction of the distance or of the dioptric power of the deformable mirror would be necessary, making the system significantly more complex. 
Especially difficult to be compensated are fast variations that might occur during the pulsed amplification process \cite{friebel:hal-01222067}.
Contrarily, the compensation achieved in the Fourier-based amplifier architecture presented in this work is passive, and thus well suited also for fast variations of $\Delta V$.

\section{From 4f-imaging to a double Fourier transform with filtering}
\label{sec:2f-4f}
\begin{figure}[t!]
\centering
  \begin{subfigure}[t]{.9 \linewidth}
    \centering
    \includegraphics[width= \linewidth]{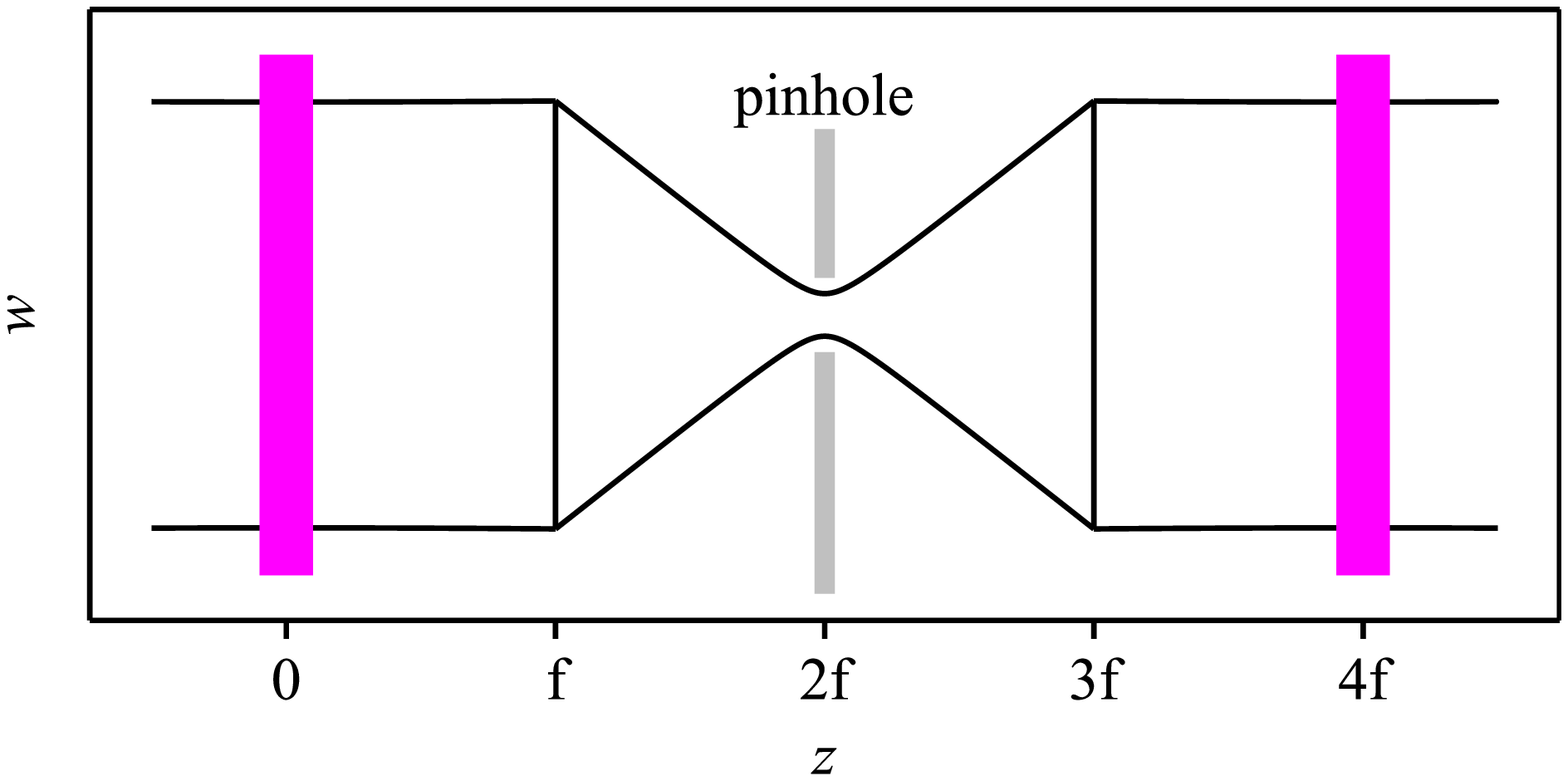} 
    \caption{} \label{fig:4f-2f-a}
  \end{subfigure}
  \begin{subfigure}[t]{.9 \linewidth}
    \centering
    \includegraphics[width= \linewidth]{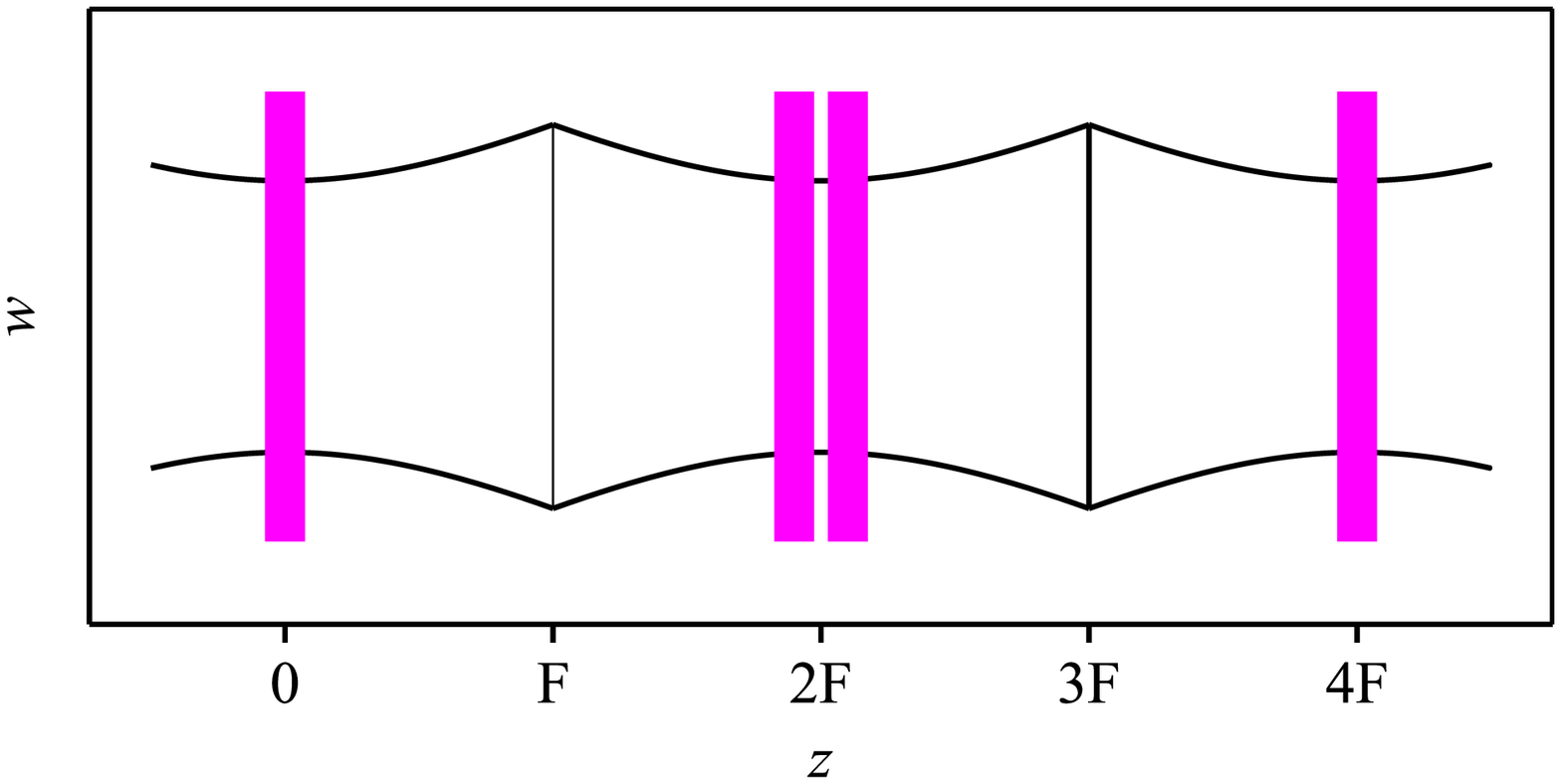} 
    \caption{} \label{fig:4f-2f-b}
  \end{subfigure}
  \caption{Schemes showing the transition from a 2-pass amplifier based   on a 4f-imaging with mode-cleaning achieved with a pinhole (a), to   a 4-pass amplifier based on a double optical Fourier transform with  mode-filtering achieved through the soft apertures of the pumped active  medium (b).  The optical Fourier transform is realized by the  sequence: free propagation of length F, focusing element with focal  length F, followed by a free propagation of length F. Same notation  as in Fig.~\ref{fig:4f-segment} is used. }
\label{fig:4f-2f}
\end{figure}
In some realizations of multi-pass amplifiers based on 4f-imaging~\cite{Hunt:78, Sullivan:91,Scott:01, rambo2016, Banerjee:15,Yu:12}, the problematic accumulations of beam distortions at each pass have been suppressed by placing tight apertures (pinholes) in the focus (center) of the 4f-imaging segments as shown in Fig.~\ref{fig:4f-2f-a}.
Mode filtering is thus achieved at the cost of output efficiency.
In these designs the 4f-propagation is split into a 2f-propagation, an aperture and a second 2f-propagation.
The 2f-propagations act as optical Fourier transforms, while the pinhole filters the beam components beyond the fundamental Gaussian mode.

These schemes indicate a way of how to modify the 4f-imaging propagation to reduce the accumulation of beam deformations: namely by using optical Fourier-transform propagations to transport the beam from pass to pass, and by using the aperture available in the active medium itself as a mode cleaner.
In this new scheme, the pinhole of Fig.~\ref{fig:4f-2f-a} is replaced by the active medium (more precisely two passes in the active media) as shown in Fig.~\ref{fig:4f-2f-b} that exhibits an intrinsic soft aperture.
Hence, the active medium acts not only as gain medium but also as mode cleaner, maximizing the gain for the fundamental mode. 
For comparison, a 4f-based 4-pass amplifier has similar aperture losses for the fundamental mode as the multi-pass amplifier of Fig.~\ref{fig:4f-2f-b}, but it does not provide mode cleaning.
Multi-pass amplifiers based on this succession of Fourier-propagations and passes at the gain medium featuring a soft aperture will be evaluated in the following sections.

\section{2-pass amplifier with Fourier transform propagations}
\label{sec:2-pass}

\begin{figure}[t!]
\centering
  \begin{subfigure}[t]{.48 \linewidth}
    \centering
    \includegraphics[width= \linewidth]{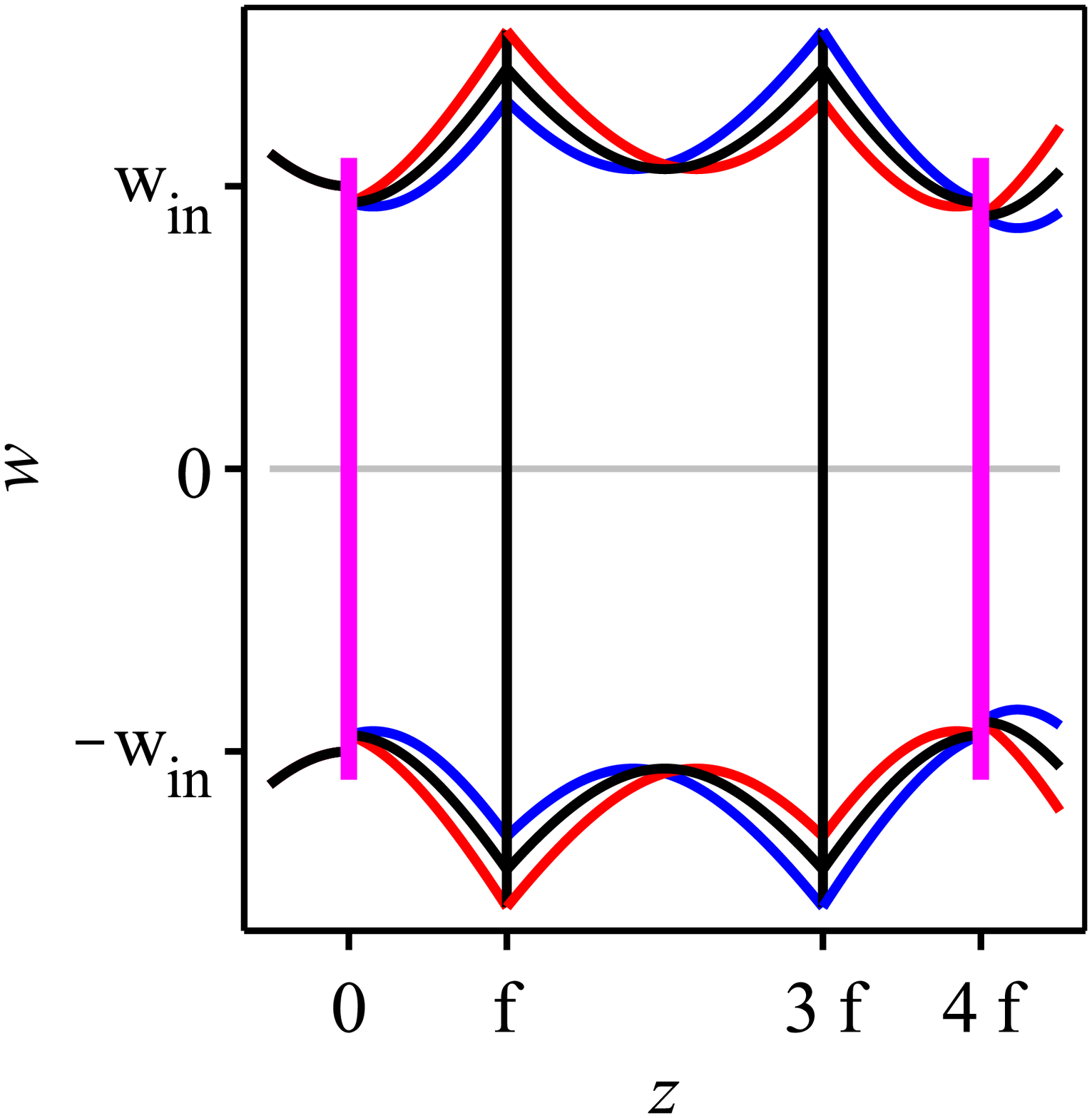} 
    \caption{} \label{Relay}
  \end{subfigure}
  \hfill
  \begin{subfigure}[t]{.48 \linewidth}
    \centering
    \includegraphics[width= \linewidth]{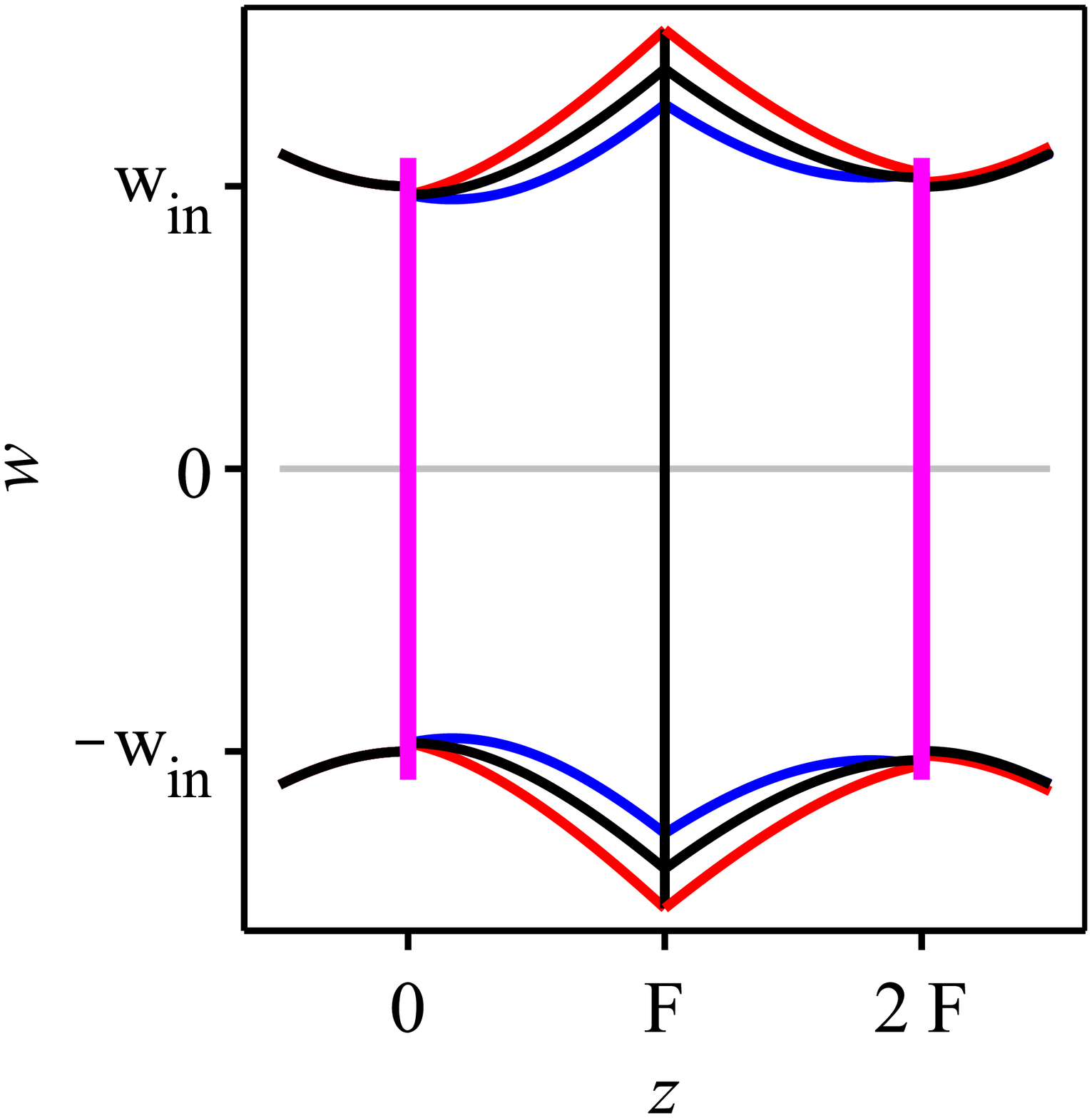} 
    \caption{} \label{Fourier}
  \end{subfigure}
\caption{Scheme and corresponding beam size evolution along the   optical axis z for a 2-pass amplifier based on 4f-imaging (a) and on an optical Fourier transform (b). 
 It was assumed that   $W=4w_\mathrm{in}$ and same notation as in Fig. \ref{fig:4f-segment}  is used. 
Similarly to Fig.~\ref{fig:4f-segment-b} the Gaussian aperture related to the finite pump region causes the discontinuity of the beam size at the gain medium.}
\label{Fig2}
\end{figure}

In this section, we investigate the properties of a Fourier-based 2-pass amplifier as sketched in Fig.~\ref{Fourier} here the beam crosses first the active medium, propagates through an optical segment acting as an optical Fourier transform, and then passes a second time the active medium.
The properties of this system are compared to the 2-pass amplifier shown in Fig.~\ref{Relay} which is based on a 4f-imaging propagation.

To quantify the sensitivity to variations of $\Delta \tilde{V}$ of the beam leaving the 2-pass amplifier we first evaluate the corresponding ABCD-matrix.
The ABCD-matrix of the 2-pass amplifier $M_\mathrm{Fourier, \, 2}$ based on the Fourier transform results from the product 
\begin{eqnarray}
  M_\mathrm{Fourier, \, 2} &= & M_\mathrm{AM}\cdot M_\mathrm{Fourier}\cdot M_\mathrm{AM}\\
  &= &\left[ \begin{array}{l l}
-F\cdot \Delta \tilde{V} & F \\ 
-\frac{F^2\cdot {\Delta \tilde{V}}^2+1}{F} & - F\cdot \Delta \tilde{V} \end{array}
  \right]\ ,
\label{eq:ABCD-matrix-2}
\end{eqnarray}
where $M_\mathrm{Fourier}$ is the ABCD-matrix of the optical Fourier transform with $A=D=0$, $B=F$, $C=-1/F$, with $F$ being the focal length of the optics (lens or curved mirror) used to realize the Fourier transform when $\Delta \tilde{V}= \Delta V +i\lambda/\pi W^2$.

By combining Eq.~(\ref{eq:ABCD-matrix-2}) with Eq.~(\ref{eq:complex-beam-evolution}) we can deduce the complex beam parameter that is replicated from the input plane (just before the first pass) to the output plane (just after the second pass) $q_\mathrm{out} = q_\mathrm{in}$.
The solution reads:
\begin{equation}
  q_\mathrm{stable}=\frac{iF}{\sqrt{1-\left( F \cdot \Delta\tilde{V} \right)^2}} \ .
  \label{eq:condition-2a}
\end{equation}
Therefore, unlike the 4f-imaging that can reproduce any beam size for any value of f, the Fourier transform can reproduce the size of the input TEM00 beam at its output only provided that the complex parameter of the input beam fulfills the condition
$q_\mathrm{in}=q_\mathrm{stable}$.
Note that in this case the output beam size equals the input beam size only when the number of passes in the active medium is even.

When designing multi-pass amplifiers the pragmatic choice for the input beam is to assume that the active medium has no dioptric power and no aperture effects ($V = \Delta V=0$ and $W=\infty$), i.e., $\Delta \tilde{V}=0$.
In this case, the stability condition given in Eq.~(\ref{eq:condition-2a}) simplifies to $q_\mathrm{stable}= i F$ so that our practical choice for the input beam parameter is 
\begin{equation}
q_\mathrm{in}= i F \ .
\label{eq:condition-1}
\end{equation}
Using Eq.~(\ref{eq:complex-beam-parameter}), this condition can be re-expressed as
\begin{equation}
  w_\mathrm{in}=\sqrt{\frac{\lambda F}{\pi}} \ .
  \label{eq:condition-2}
\end{equation}
It is evident from this equation that the focal length $F$ needed to realize the Fourier transform reproducing the size of the beam from active medium to active medium depends on the desired beam size at the active medium.

The benefits of the Fourier transform can be evinced by following the beam propagation shown in Fig.~\ref{Fourier}.
The beam size of the input beam $w_\mathrm{in,\, 1}$ is reduced by the the soft aperture with $W\neq \infty$ of the active medium, so that $w_\mathrm{out, \, 1}<w_\mathrm{in, \, 1}$.
The smaller beam is, however, converted by the Fourier transform into a larger beam $w_\mathrm{in, \, 2}>w_\mathrm{in, \, 1}$ (see~Fig.~\ref{Fourier}).
At the second pass through the active medium, this larger beam is reduced by the aperture: $w_\mathrm{out, \, 2}<w_\mathrm{in, \, 2}$ so that the resulting beam has roughly the size of the input beam $w_\mathrm{out, \, 2}\approx w_\mathrm{in,\, 1}$.
There is thus a counteracting interplay between the reduction of the beam size caused by the aperture, and the increase of the beam size related with the Fourier transform, that stabilizes the beam size close to the value expressed by Eq.~(\ref{eq:condition-2}) so that  similar beam sizes are reproduced at each pass.
Because of this compensation, variations of the width $W$ of the soft aperture do not significantly affect the output beam size so that a precise knowledge of the aperture effect is not needed when designing  Fourier-based multi-pass systems.

For an analogous compensation mechanism, the Fourier-based 2-pass amplifiers exhibit output beam sizes and divergences that are insensitive to variations of  $\Delta V$ (see Fig.~\ref{Fourier}).
Indeed, an active medium with focusing dioptric power turns the entering collimated beam into a converging beam, but the following Fourier transform converts this converging beam into a diverging beam.
At the second pass, the focusing dioptric power of the active medium essentially cancels the divergence of the impinging beam.
In such a way, the beam leaving the 2-pass amplifier has similar size and divergence as the collimated input beam.
An analogous compensation occurs for an active medium with defocussing dioptric power.

Summarizing, there is a counter-action between the Fourier transform and the active medium that results in output beam characteristics only weakly dependent on variations of $\Delta \tilde{V}$.
The reduced sensitivity to variations of $\Delta V$ of the Fourier-based compared to the 4f-based amplifier can be clearly appreciated by comparing the divergences of the output beams in Figs.~\ref{Relay} and ~\ref{Fourier}.
The 4f-imaging leads to an accumulation of the lens effects from pass to pass, while the Fourier transform leads to a substantial compensation of these lens effects.
This compensation applies also for other deformations of the phase front occurring at the active medium beyond the spherical distortion already accounted in $V$.

The complex parameter of the outgoing beam after propagation along the two pass amplifier can be calculated from Eq.~(\ref{eq:complex-beam-evolution}) and the ABCD-matrix elements of Eq.~(\ref{eq:ABCD-matrix-2}).
By assuming an input beam parameter as given in Eq.~(\ref{eq:condition-1}),  we find that the complex parameter of the output beam for small variations of $\Delta \tilde{V}$ takes the form
\begin{equation}
  q_\mathrm{out, \, 2} = i F + i F^3 {\Delta \tilde{V}}^2 -
  F^4 {\Delta \tilde{V}}^3-F^6 {\Delta \tilde{V}}^5 + \cdots
  \label{eq:q-2}
\end{equation}
after a Taylor expansion around $q_\mathrm{in}=iF$.

Remarkably, in this case $q_\mathrm{out, \, 2}$ depends only quadratically on variations of the complex-valued dioptric power $\Delta \tilde{V}$.
This is in sharp contrast to the linear dependence observed in Eq.~(\ref{eq:Taylor-4f}) for the amplifier based on 4f-imaging.  
In other words, the output beam for the 2-pass Fourier-based amplifier in first order does not depend on $\Delta \tilde{V}$, so that the output beam characteristics such as phase front curvature (given by radius $R_\mathrm{out, \, 2}$) and beam size $w_\mathrm{out, \, 2}$ are very insensitive to small variations of the dioptric power and aperture effects of the active medium.

\begin{figure}[tb!]
\centering
  \begin{subfigure}[t]{.45 \linewidth}
    \centering
    \includegraphics[width= \linewidth]{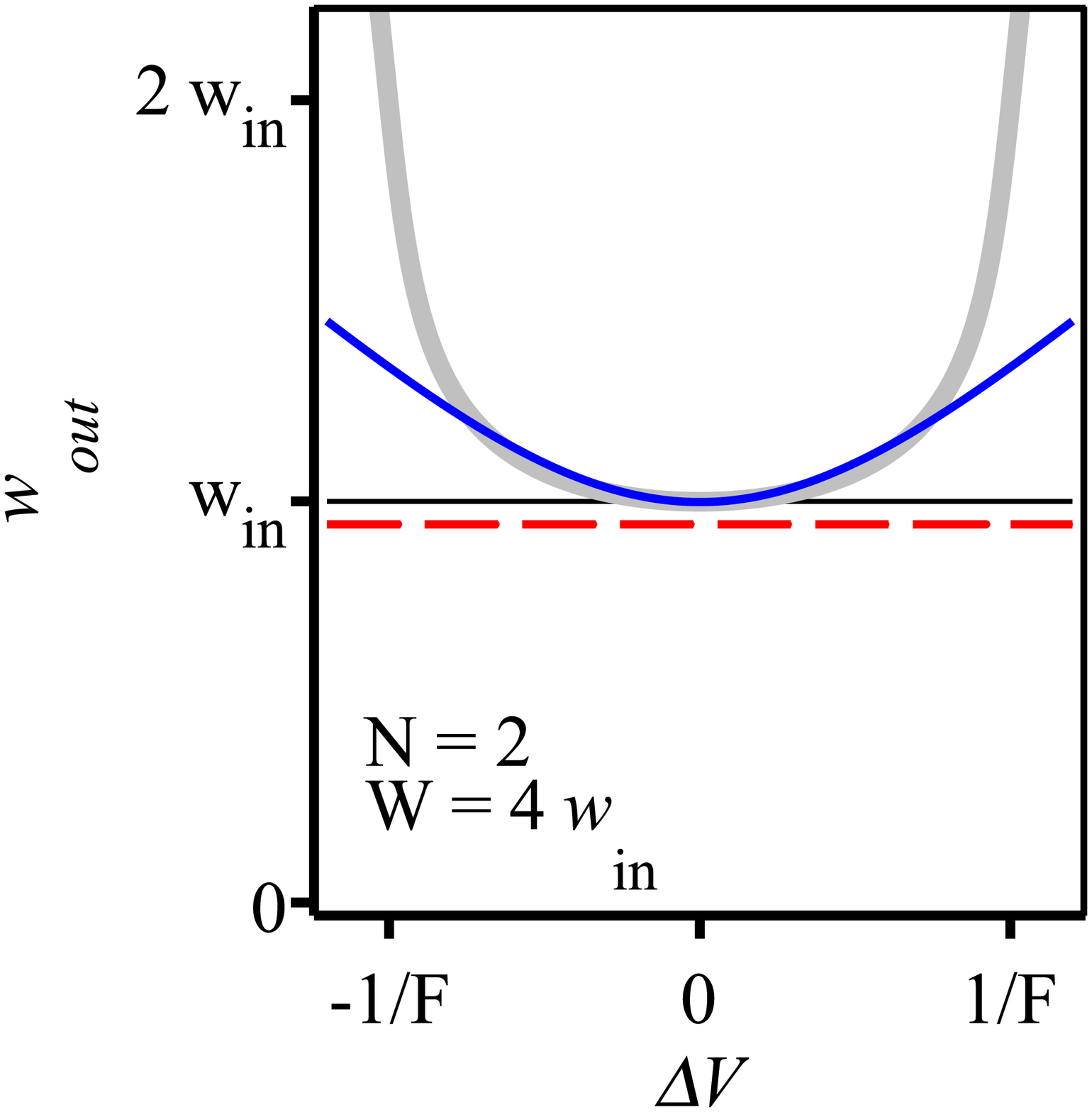} 
    \caption{} \label{Waist1}
  \end{subfigure}
  \hfill
  \begin{subfigure}[t]{.45 \linewidth}
    \centering
    \includegraphics[width= \linewidth]{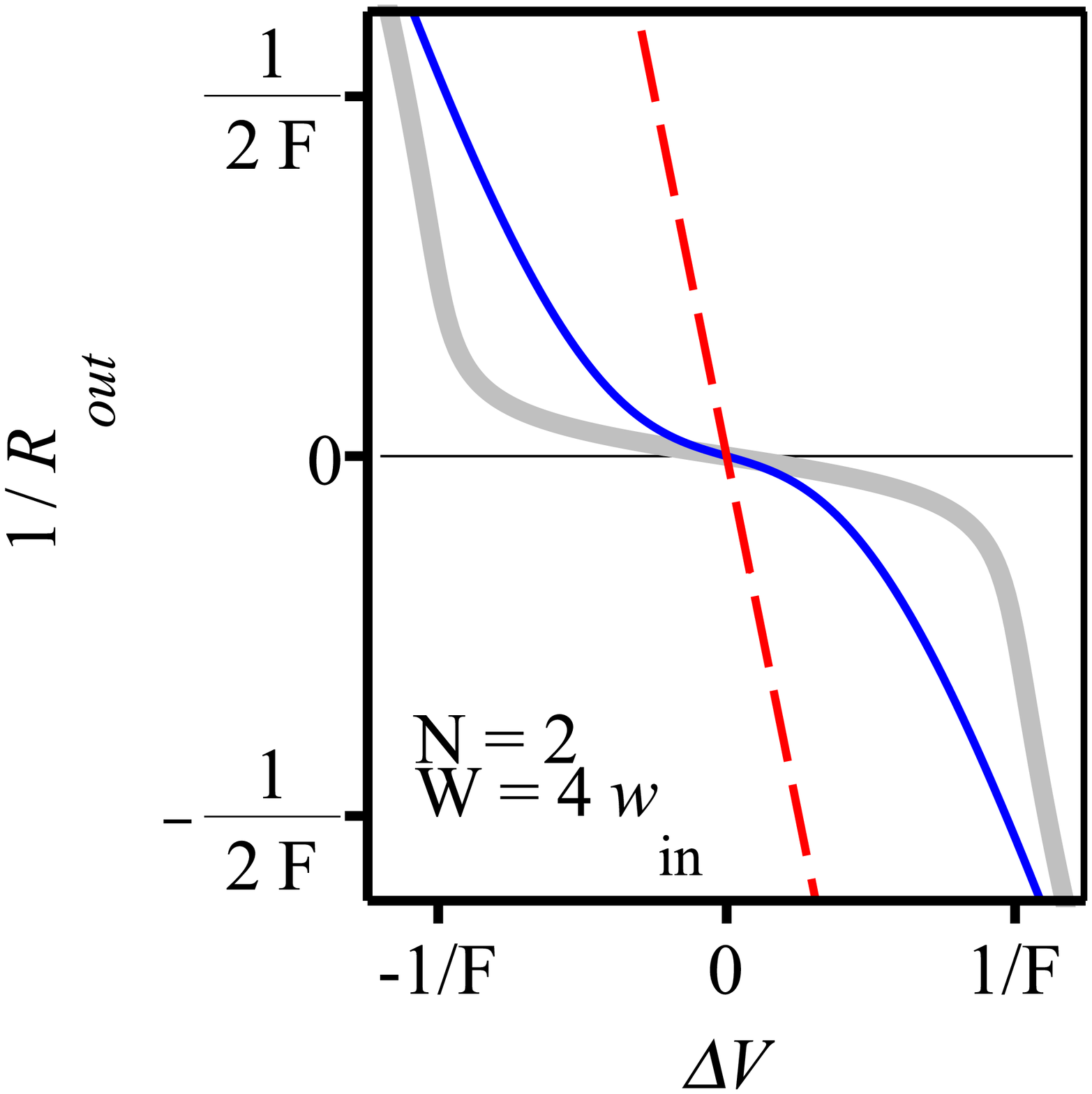} 
    \caption{} \label{Curvature1}
  \end{subfigure}
    \begin{subfigure}[t]{.45 \linewidth}
    \centering
    \includegraphics[width= \linewidth]{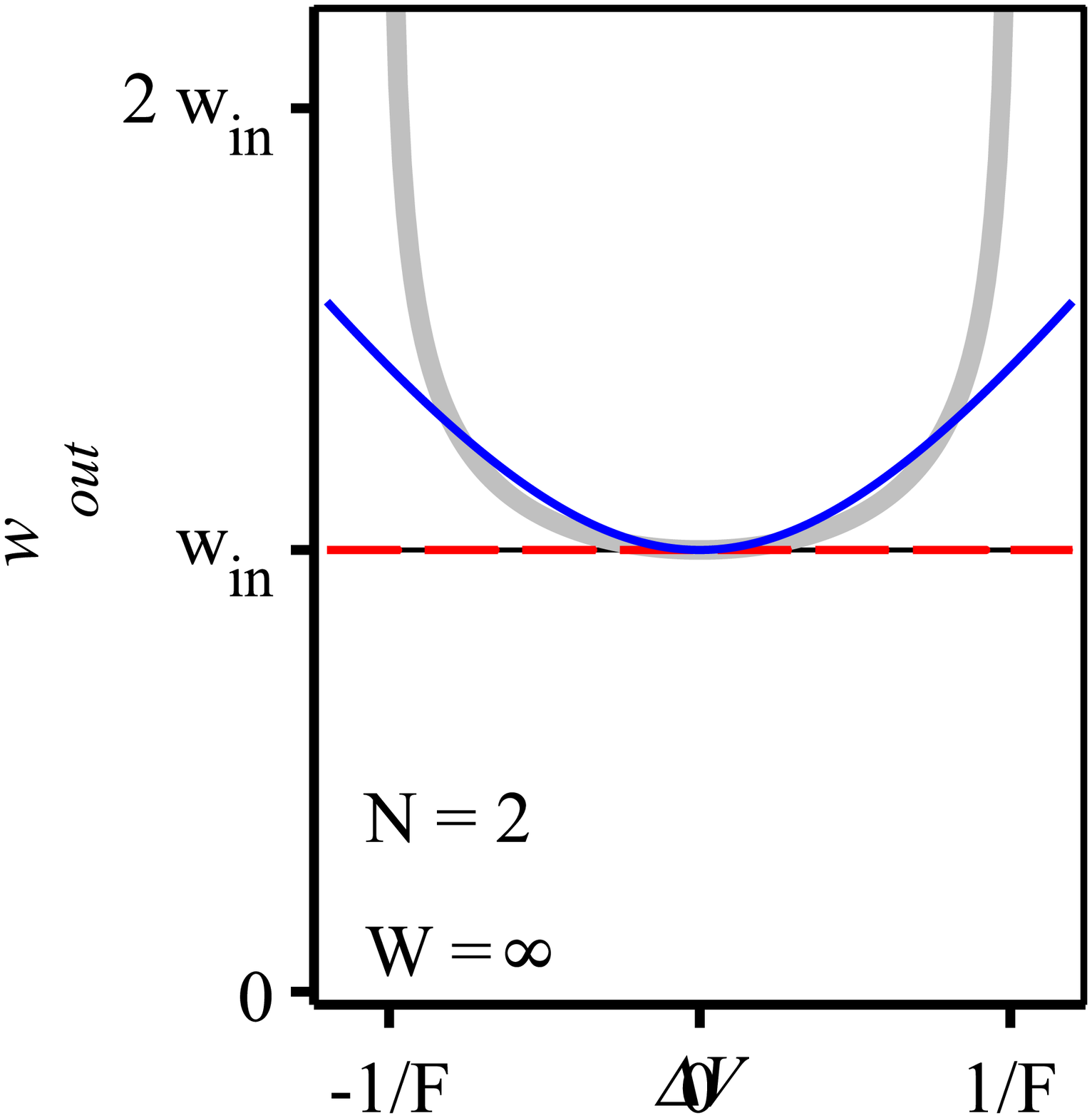} 
    \caption{} \label{Waist1_40}
  \end{subfigure}
  \hfill
  \begin{subfigure}[t]{.45 \linewidth}
    \centering
    \includegraphics[width= \linewidth]{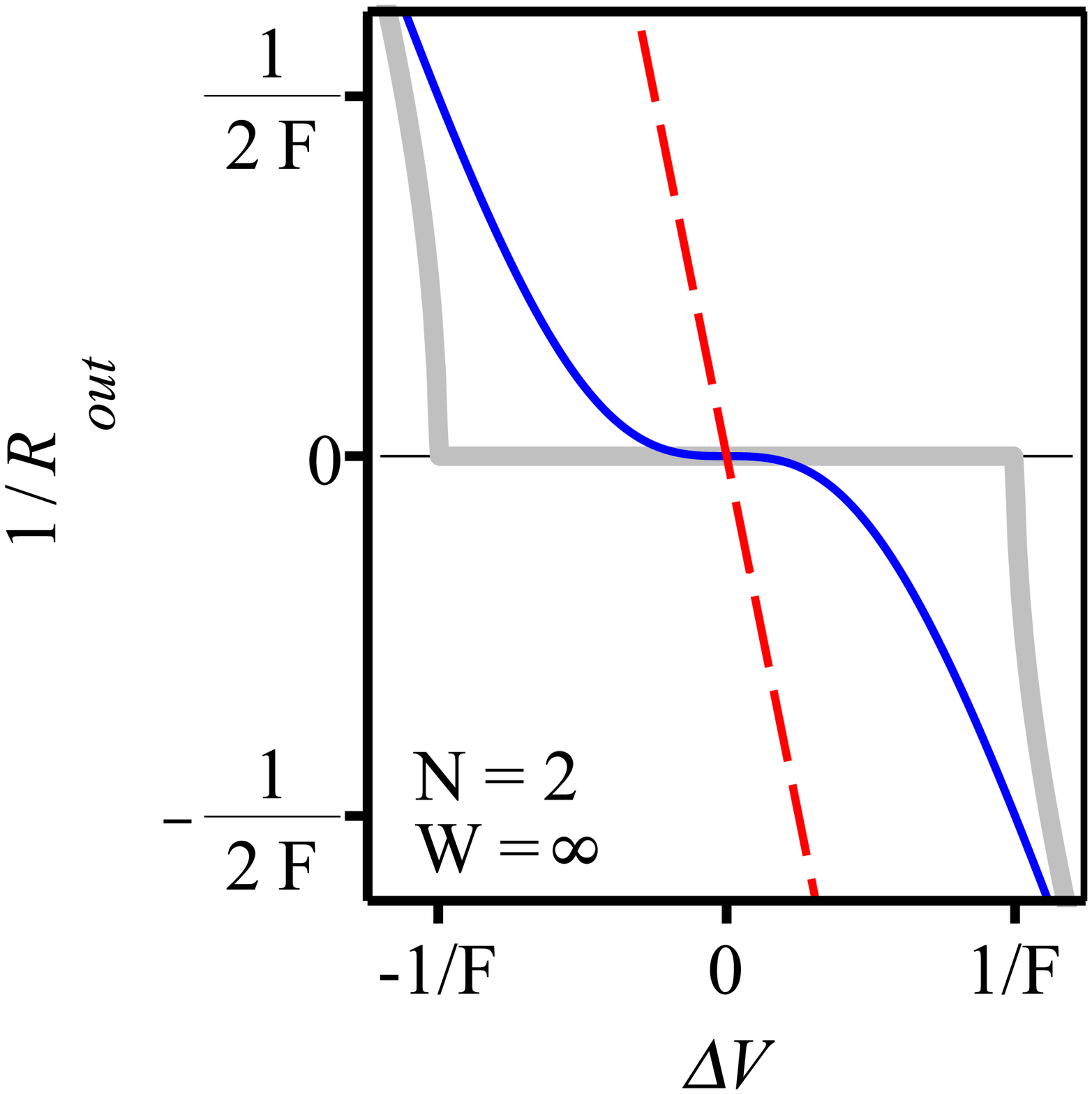} 
    \caption{} \label{Curvature1_40}
  \end{subfigure}
  \caption{Size $w_\mathrm{out, \, 2}$ and phase front curvature     $R^{-1}_\mathrm{out, \, 2}$ of the output beam leaving  the 2-pass amplifiers of Fig.~\ref{Fig2} for variations of the    active medium dioptric power $\Delta V$. 
The top row assumes $W=4w_\mathrm{in}$, the bottom one $W=\infty$. 
The blue curves represent the results for the Fourier-based amplifier of Fig.~\ref{Fourier}, while the dashed red  curves represent the results for the 4f-based amplifier of Fig.~\ref{Relay}. 
In both cases it was assumed that $q_\mathrm{in}=q_\mathrm{stable}$. For comparison in gray are    given $w_\mathrm{stable}(\Delta V)$ and $R^{-1}_\mathrm{stable}(\Delta V)$ corresponding to the complex beam parameter given in Eq.~(\ref{eq:condition-2a}).  }
\label{Fig3}
\end{figure}

The beam size $w_\mathrm{out, \, 2}$ and the phase front curvature $R_\mathrm{out, \, 2}$ obtained from $q_\mathrm{out, \, 2}$ are shown in blue in Figs.~\ref{Waist1} and \ref{Curvature1} respectively, for variations of the dioptric power $\Delta V$ and aperture size of $W=4w_\mathrm{in}$.
Similar plots are given in Figs.~\ref{Waist1_40} and \ref{Curvature1_40} for $W=\infty$ to highlight the role of the aperture.
When comparing these blue curves with the red dashed curves, that represent the output beam characteristics for the 4f-based 2-pass amplifier, it becomes evident that for a 4f-based amplifier the output beam size strongly depends on $W$ and that the output  divergence strongly depends on $\Delta V$.
Oppositely, the Fourier-based multi-pass amplifier has output characteristics insensitive to small variations of $\Delta V$ and $W$.

Assuming no aperture effects ($W=\infty$), the Taylor expansions of the output beam characteristics take the form
\begin{eqnarray}
  \frac{w_\mathrm{out, \, 2}}{w_\mathrm{in}} &\hspace{-3mm}= &\hspace{-3mm}1+ \frac{1}{2}F^2\Delta V^2-\frac{1}{8}F^4\Delta V^4+ \cdots 
  \label{eq:2-pass-w}
  \\
R^{-1}_\mathrm{out, \, 2}&\hspace{-3mm}=&\hspace{-3mm}-{F}^{2}{\Delta V}^{3}+{F}^{4}{\Delta V}^{5}-{F}^{6}{\Delta V}^{7}+{F
}^{8}{\Delta V}^{9}+ \cdots 
  \label{eq:2-pass-R}
\end{eqnarray}
with $w_\mathrm{in}$ satisfying Eq.~(\ref{eq:condition-2}).
The insensitivity of the output beam properties to small variations of the active medium dioptric power is manifested graphically in Fig.~\ref{Fig3} by the flat behavior of the blue curves around $\Delta V=0$. 
Mathematically it shows in the absence of a linear term in $\Delta V$ in Eq.~(\ref{eq:2-pass-w}), and by the absence of both liner and quadratic terms in $\Delta V$ in Eq.~(\ref{eq:2-pass-R}).

It is also interesting to investigate how the Gaussian aperture of the active medium affects the output beam size.
Assuming that $V=\Delta V=0$, we find that the output beam size takes the simple form
\begin{equation}
  \frac{w_\mathrm{out, \, 2}}{w_\mathrm{in}} =1  +\frac{w_\mathrm{in}^4}{2W^4}
  -\frac{w_\mathrm{in}^8}{8W^8}+\frac{w_\mathrm{in}^{12}}{16W^{12}}+\cdots \ .
  \label{eq:2-pass-w2}
\end{equation}
Hence,  the relative change of the output beam size caused by the aperture is in leading order dependent on $w^4_\mathrm{in}/2W^4$ compared to the $w^2_\mathrm{in}/W^2$ for the 4f-based 2-pass amplifier (see Eq.~(\ref{eq:4f-R-b})).
In other words, aperture effects minimally impact the size of the output beam of a Fourier-based multi-pass amplifier as long as $w_\mathrm{in}/W \ll 1$.

The dependence of the phase front radius $R^{-1}_\mathrm{out, \, 2}$ on aperture effects has also been investigated.
Similar to the 4f-based amplifier, also in this case the phase front radius (divergence) of the beam leaving the amplifier, for $V=0$, does not depend on the aperture size $W$.

\section{4-pass  amplifiers with Fourier transform propagations}
\label{sec:4-pass}

\begin{figure}[htbp]
\centering
  \begin{subfigure}[t]{.95 \linewidth}
    \centering
    \includegraphics[width= \linewidth]{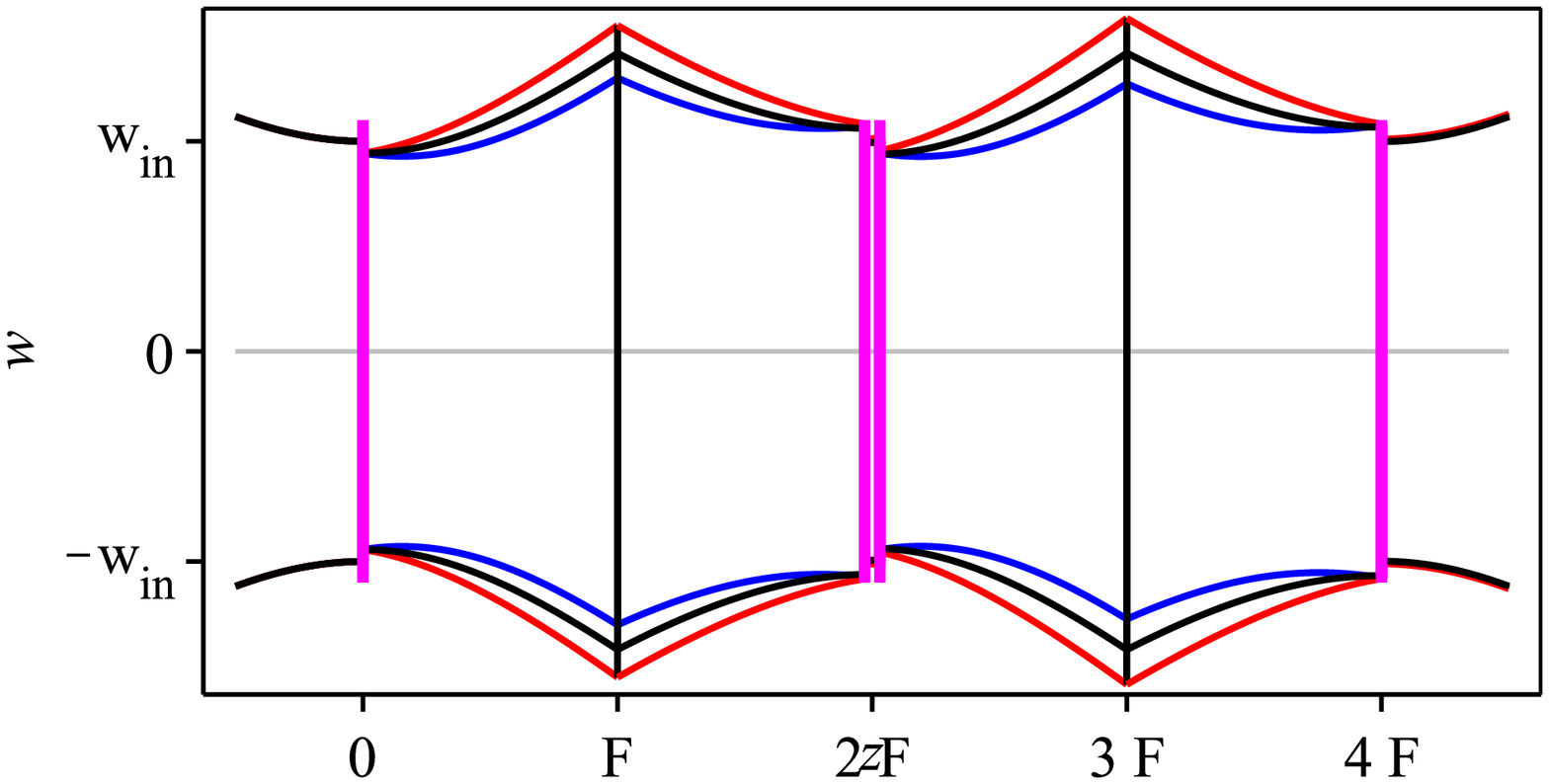} 
    \caption{} \label{Fig4}
  \end{subfigure}
  \hfill
  \begin{subfigure}[t]{.95 \linewidth}
    \centering
    \includegraphics[width= \linewidth]{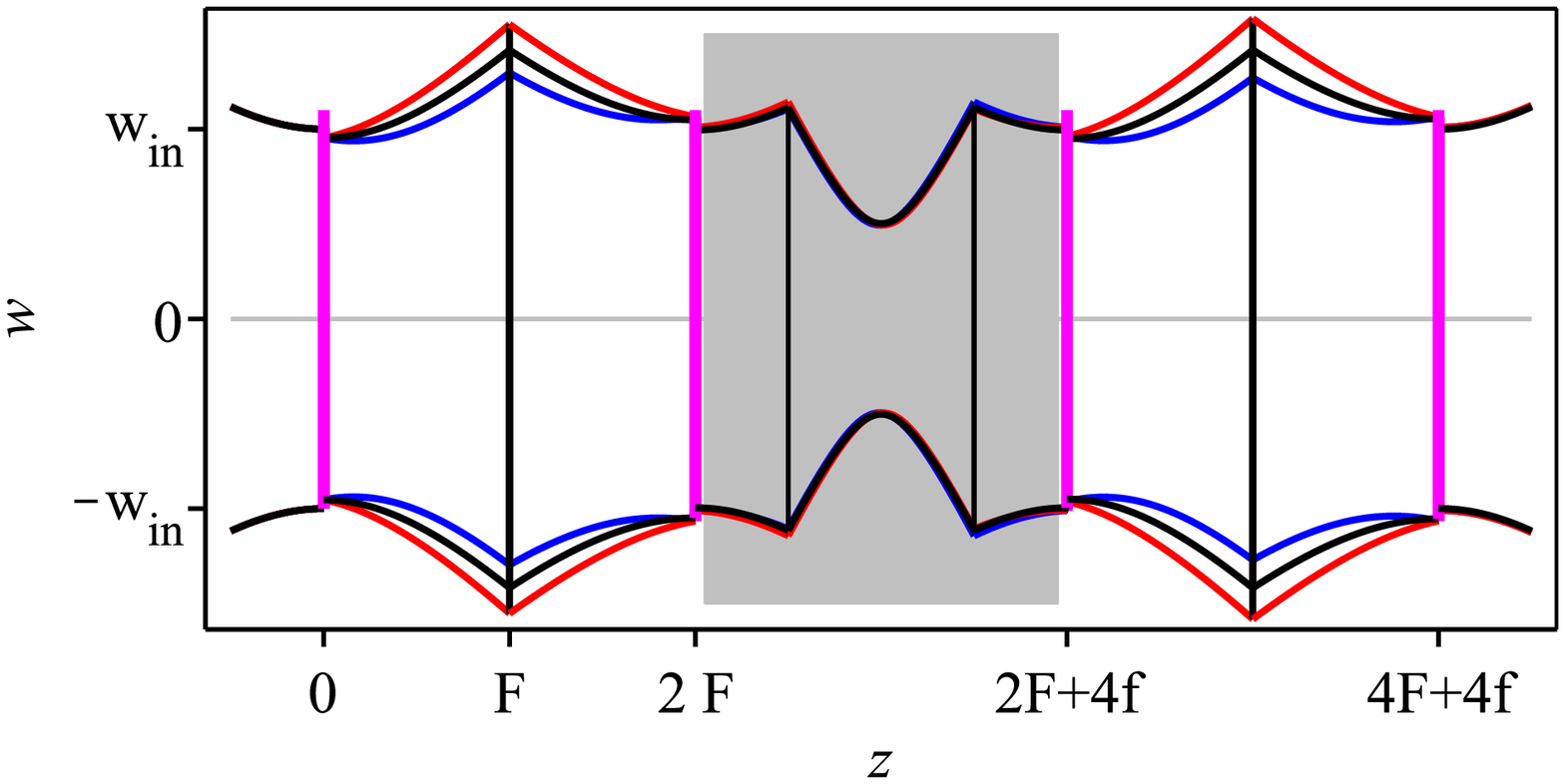} 
    \caption{} \label{Fig5}
  \end{subfigure}
  \caption{Two types of 4-pass amplifiers both based on two optical Fourier transforms.  
In (a) there is physically short free propagation  between the second and third pass.  
In (b) a 4f-imaging is used to obtain a propagation with zero effective length between the second and third pass. 
Same notation as in Fig.~\ref{fig:4f-segment} is used. The gray shaded region indicates the 4f-imaging segment.
The discontinuity of the beam size due to the aperture at the active medium is visible.}
\label{fig:4-pass-Fourier}
\end{figure}

In this section, we consider the 4-pass amplifier based on optical Fourier transforms obtained by concatenating two of the optical segments of Fig.~\ref{Fourier}.
The resulting propagation follow thus the sequence
\begin{equation*}
AM-Fourier-AM-AM-Fourier-AM,
\end{equation*}
as sketched in Fig.~\ref{Fig4}. 
We assumed here that the propagation between the second and the third pass on the active medium  is so short that it can be neglected.
This can either be realized with a propagation that is physically short compared to the Rayleigh length of the beam (see Fig.~\ref{Fig4}), or as sketched in Fig.~\ref{Fig5}, by inserting a 4f-imaging  between the second and the third pass that per definition has a vanishing effective optical length ($B=0$ for the ABCD-matrix of a 4f-imaging propagation).
In this case, the propagation in the 4-pass amplifier follows the sequence
\begin{equation*}
AM-Fourier-AM-4f-AM-Fourier-AM.
\end{equation*}

\begin{figure}[htbp]
\centering
  \begin{subfigure}[t]{.45 \linewidth}
    \centering
    \includegraphics[width= \linewidth]{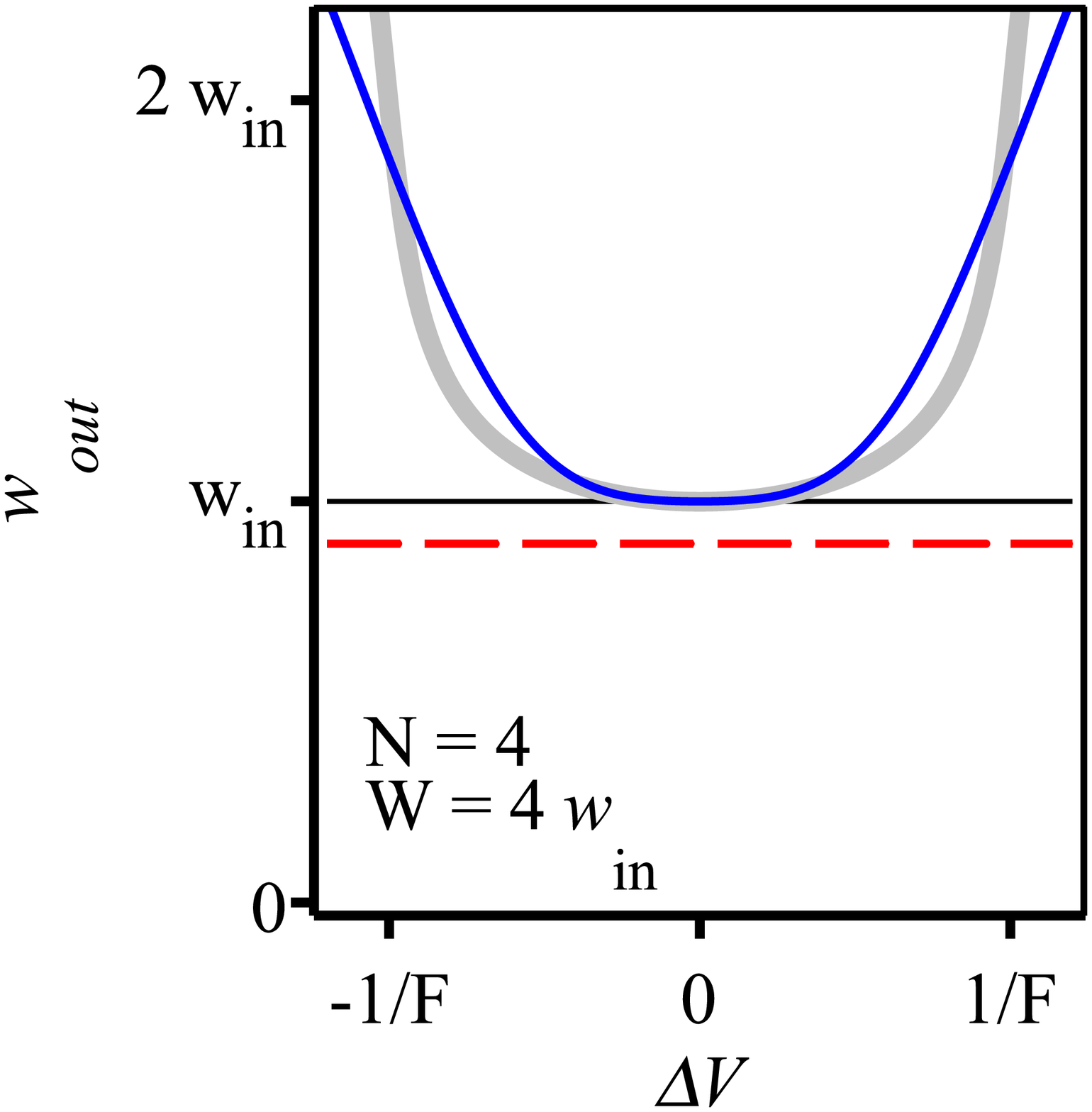} 
    \caption{} \label{Waist2}
  \end{subfigure}
  \hfill
  \begin{subfigure}[t]{.45 \linewidth}
    \centering
    \includegraphics[width= \linewidth]{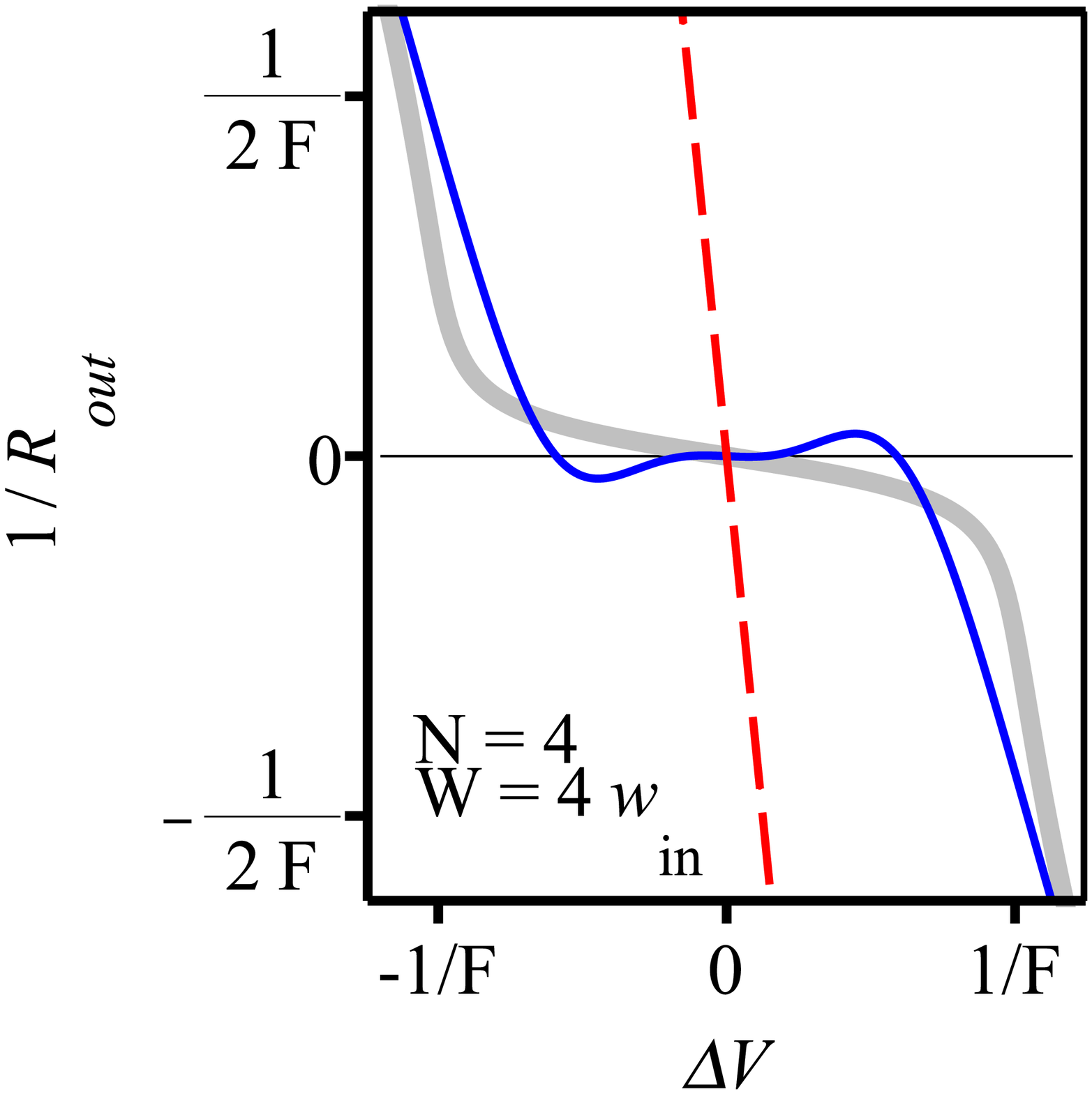} 
    \caption{} \label{Curvature2}
  \end{subfigure}
    \begin{subfigure}[t]{.45 \linewidth}
    \centering
    \includegraphics[width= \linewidth]{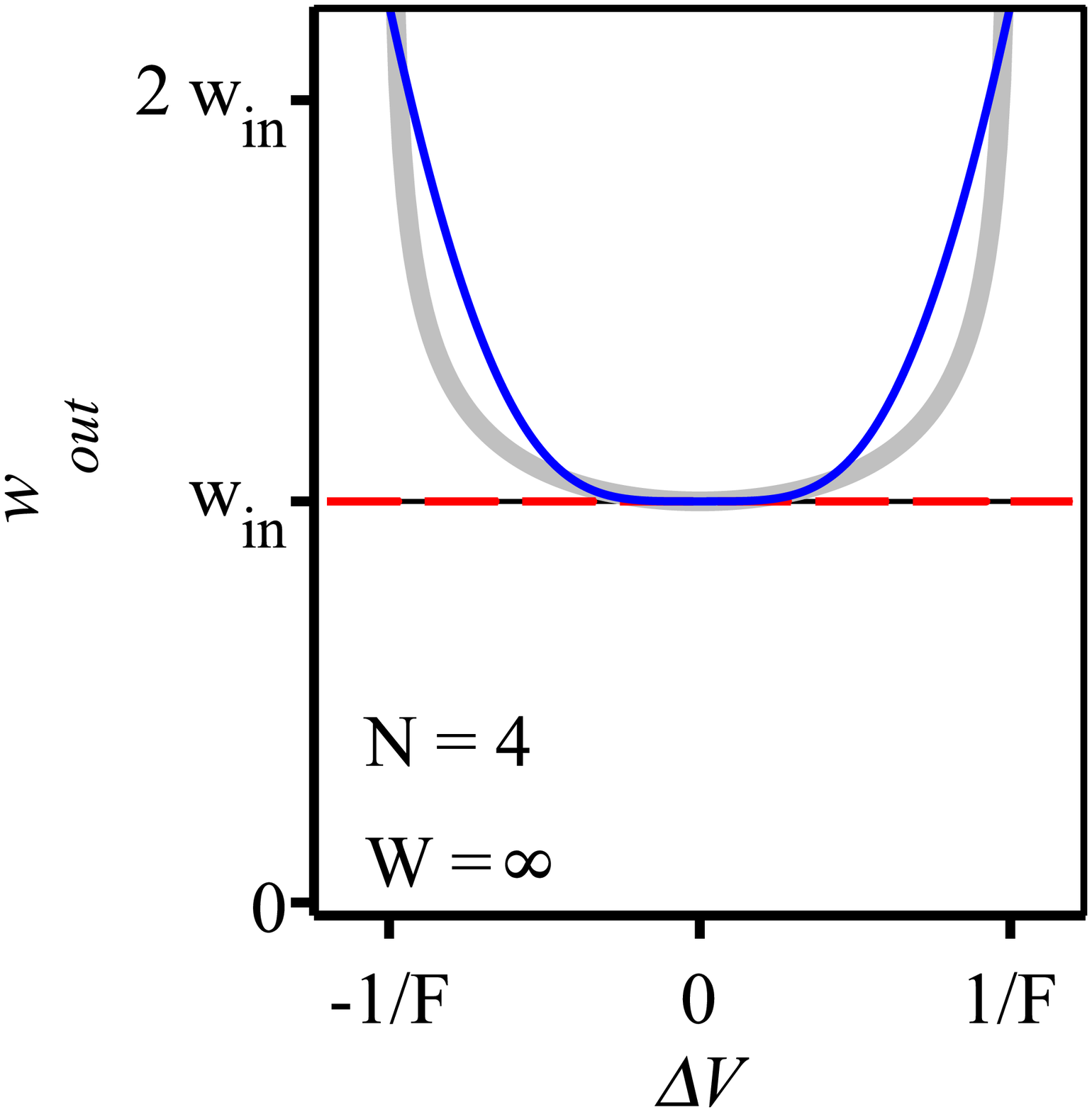} 
    \caption{} \label{Waist2_40}
  \end{subfigure}
  \hfill
  \begin{subfigure}[t]{.45 \linewidth}
    \centering
    \includegraphics[width= \linewidth]{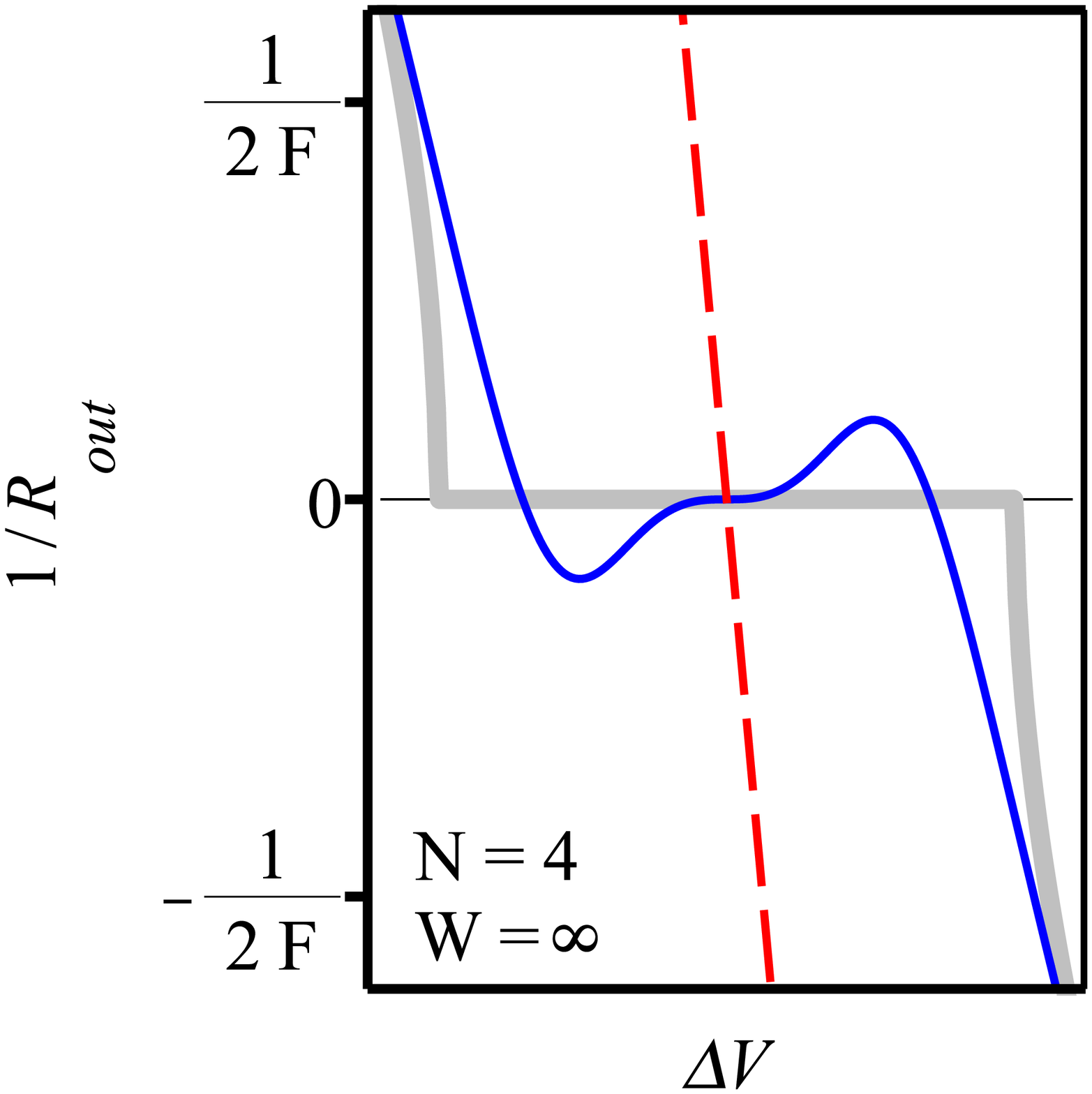} 
    \caption{} \label{Curvature2_40}
  \end{subfigure}
\caption{Similar to Fig.~\ref{Fig3} but for  amplifiers with 4 passes. 
The blue curves represent the output beam characteristics of both layouts of Fig.~\ref{fig:4-pass-Fourier}.
The red dashed curves of the output output characteristics of the 4f-amplifier design.}
\label{Fig6}
\end{figure}

The complex parameter of the output beam can be obtained following a similar procedure as in the previous section and can be approximated by a Taylor expansion in $\Delta \tilde{V}$ as
\begin{equation}
  q_\mathrm{out, \, 4} =
  i F + 2 F^4 {\Delta \tilde{V}}^3 - 4 i F^5 {\Delta \tilde{V}}^4 - 4 {F}^6 {\Delta \tilde{V}}^5+\cdots \ .
  \label{eq:4-pass-q}
\end{equation}
Notably in this equation, there are no linear and no quadratic terms in $\Delta \tilde{V}$.
Indeed, the lowest order affecting the output beam parameter depends on the third power of $\Delta \tilde{V}$.
The output beam parameters are thus even less sensitive to small variations of the thermal lens and aperture effects compared to the 2-pass Fourier-based amplifiers (compare Eq.~(\ref{eq:4-pass-q}) to Eq.~(\ref{eq:q-2})).

The size and divergence of the output beam for  $\Delta V$ variations can be obtained by combining Eq.~(\ref{eq:4-pass-q}) with Eq.~(\ref{eq:complex-beam-parameter}).
Assuming for simplicity that  $W=\infty$ we find
\begin{eqnarray}
\frac{w_\mathrm{out, \, 4}}{w_\mathrm{in}} &\!\!\!=&\!\!\!  1+ 2{F}^{4}{\Delta V}^{4}-2{F}^{8}{\Delta V}^{8}+ \cdots \\
R^{-1}_\mathrm{out, \, 4}&\!\!\!=&\!\!\! 2\,{F}^{2}{\Delta V}^{3}-4\,{F}^{4}{\Delta V}^{5}-8\,{F}^{6}{\Delta V}^{7}+ \cdots \ .
\end{eqnarray}

The insensitivity of $w_\mathrm{out, \, 4}$ to variations of the active medium dioptric power around $V=0$ roots from its $\Delta V^4$-dependence, to be compared to the $\Delta V^3$-dependence of the 2-pass Fourier-based amplifier, and the $\Delta V$-dependence of the 4f-based amplifiers.
Differently, the decreased sensitivity of $R^{-1}_\mathrm{out, \, 4}$ relative to $R^{-1}_\mathrm{out, \, 2}$ to small variations of the active medium dioptric power roots in a better cancellation between the $\Delta V^3$ and the $\Delta V^4$ terms.

A plot of $w_\mathrm{out, \, 4}$ and $R^{-1}_\mathrm{out, \, 4}$ for variations of $\Delta V$ is shown in blue in Fig.~\ref{Fig6} and compared with the results given in dashed red for a 4f-based amplifier with 4 passes and same beam size.
For the Fourier-based amplifier the ``stability region'', where the impact of $\Delta V$ variations to the output beam characteristics is small, increases with the number of passes while it decreases for 4f-based amplifiers (compare Fig.~\ref{Fig6} with Fig.~\ref{Fig3}).

\section{Multi-pass amplifier with large number of passes }
\label{sec:many-passes}

Following the reasoning of the above sections, an amplifier with large number of passes ($N=8,12,16,\cdots$) inheriting the same insensitivity to variations of the active medium lens and aperture effects can be realized by  concatenating several of the 4-pass segments presented in Fig.~\ref{fig:4-pass-Fourier}.

\begin{figure}[tp!]
\centering
\includegraphics[width=0.9 \linewidth]{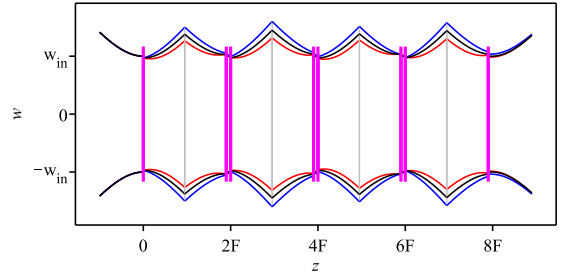}
\caption{8-pass amplifier realized by concatenating two 4-pass   amplifiers of Fig.~\ref{Fig4}. Same notation as in Fig. 1 has   been used.}
\label{fig:multy-pass-2f-only}
\end{figure}
\begin{figure}[t!]
\centering
  \begin{subfigure}[t]{.45 \linewidth}
    \centering
    \includegraphics[width= \linewidth]{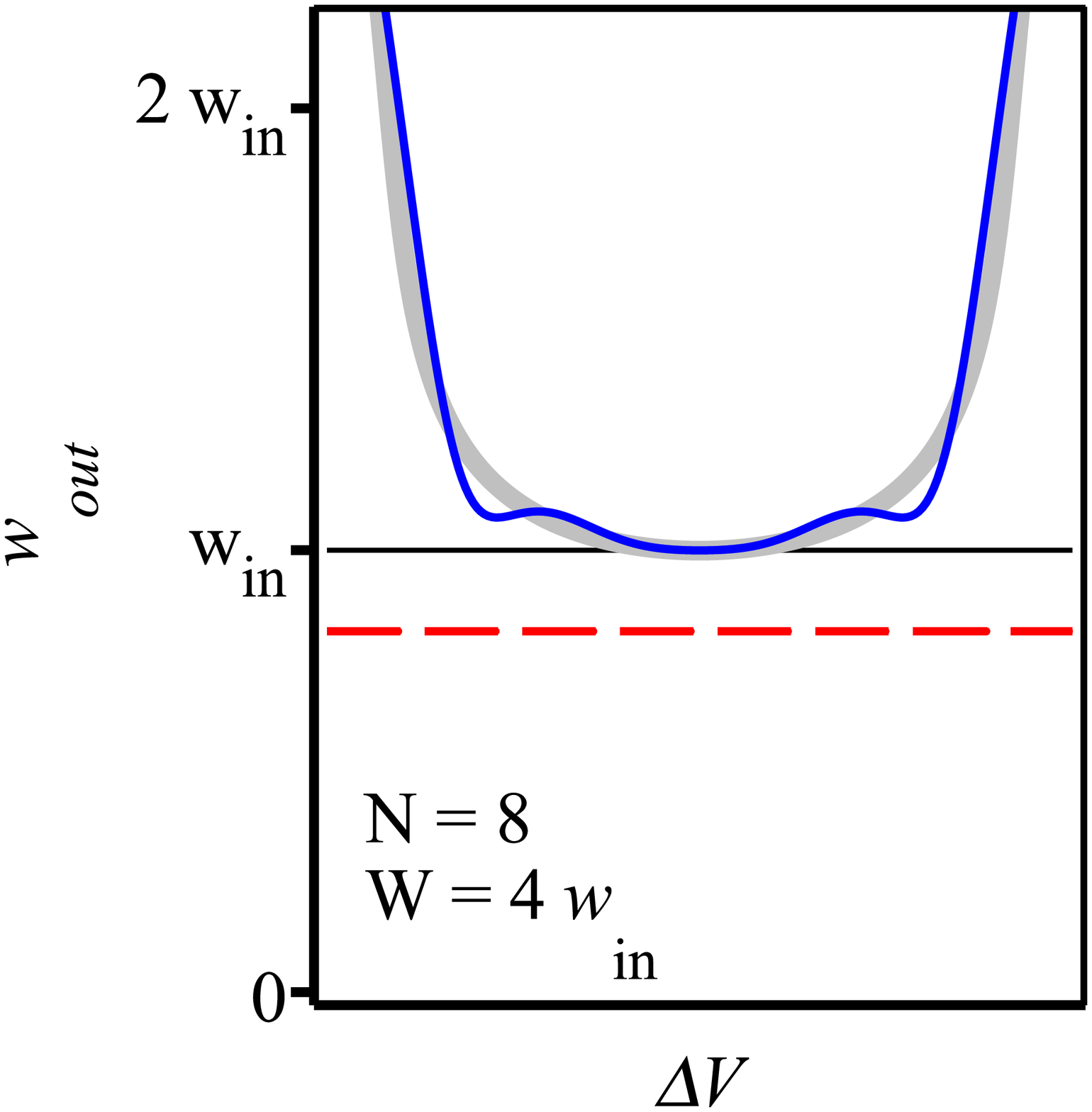} 
    \caption{} \label{Waist3}
  \end{subfigure}
  \hfill
  \begin{subfigure}[t]{.45 \linewidth}
    \centering
    \includegraphics[width= \linewidth]{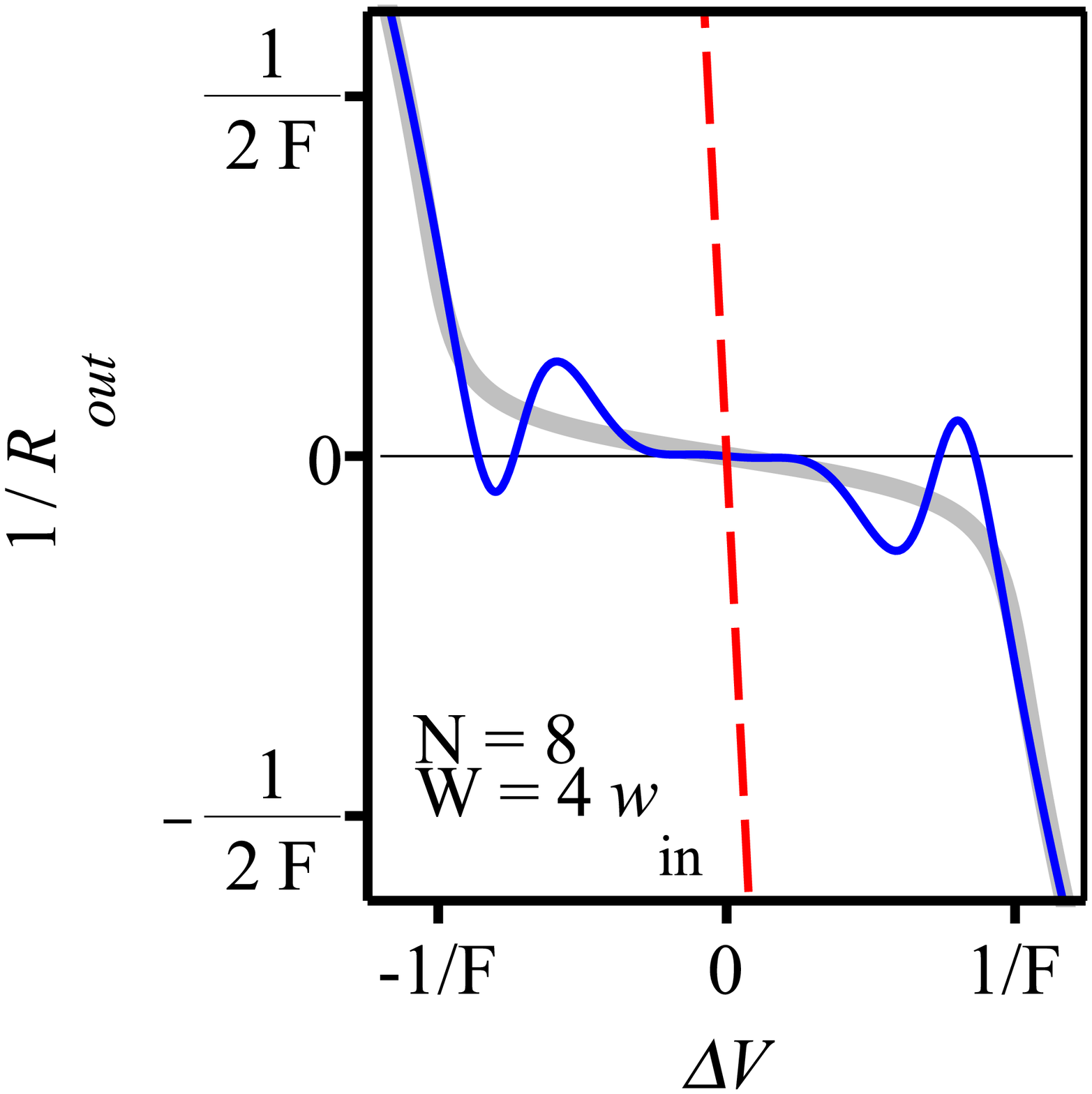} 
    \caption{} \label{Curvature3}
  \end{subfigure}
    \begin{subfigure}[t]{.45 \linewidth}
    \centering
    \includegraphics[width= \linewidth]{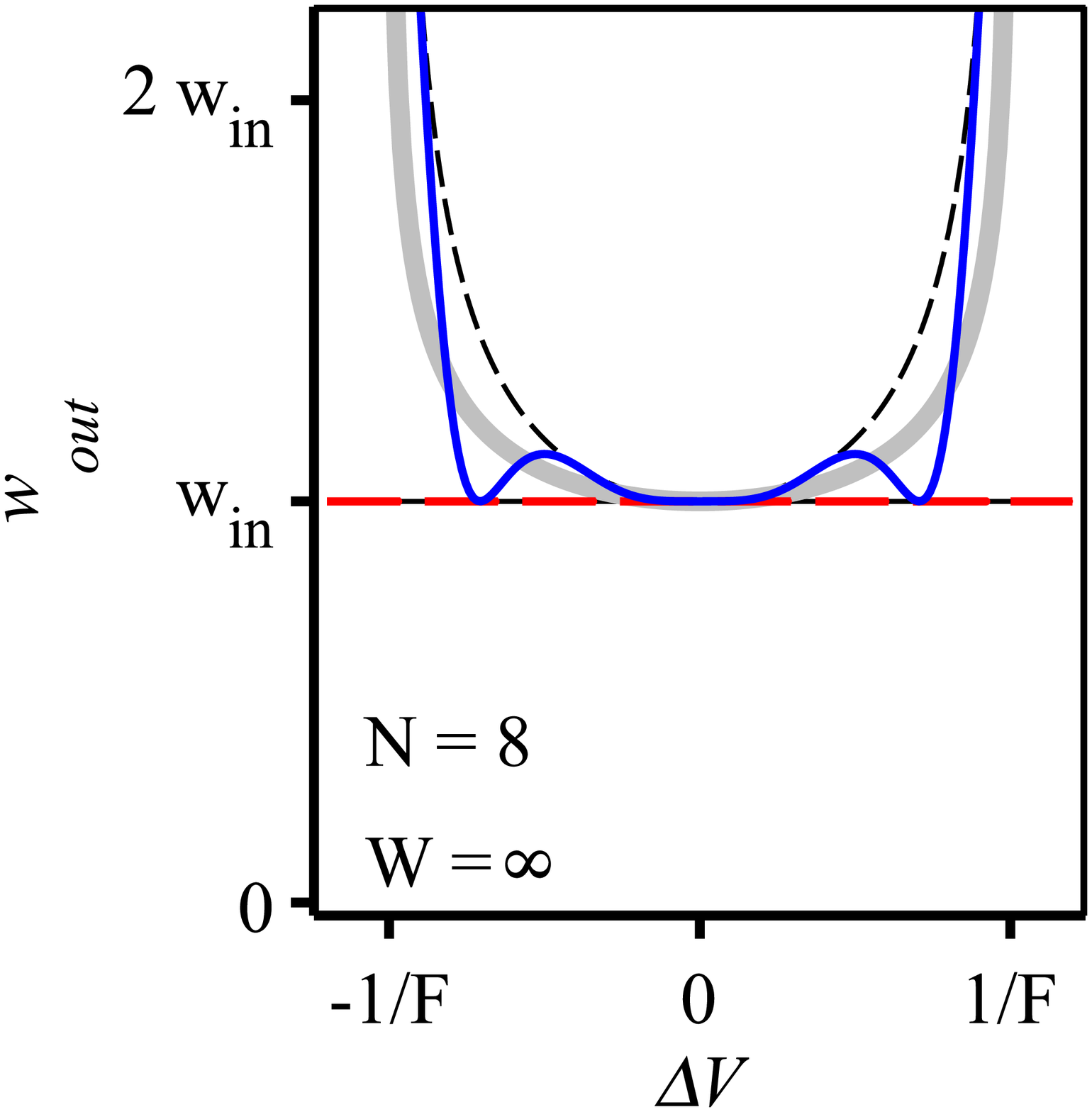} 
    \caption{} \label{Waist3_40}
  \end{subfigure}
  \hfill
  \begin{subfigure}[t]{.45 \linewidth}
    \centering
    \includegraphics[width= \linewidth]{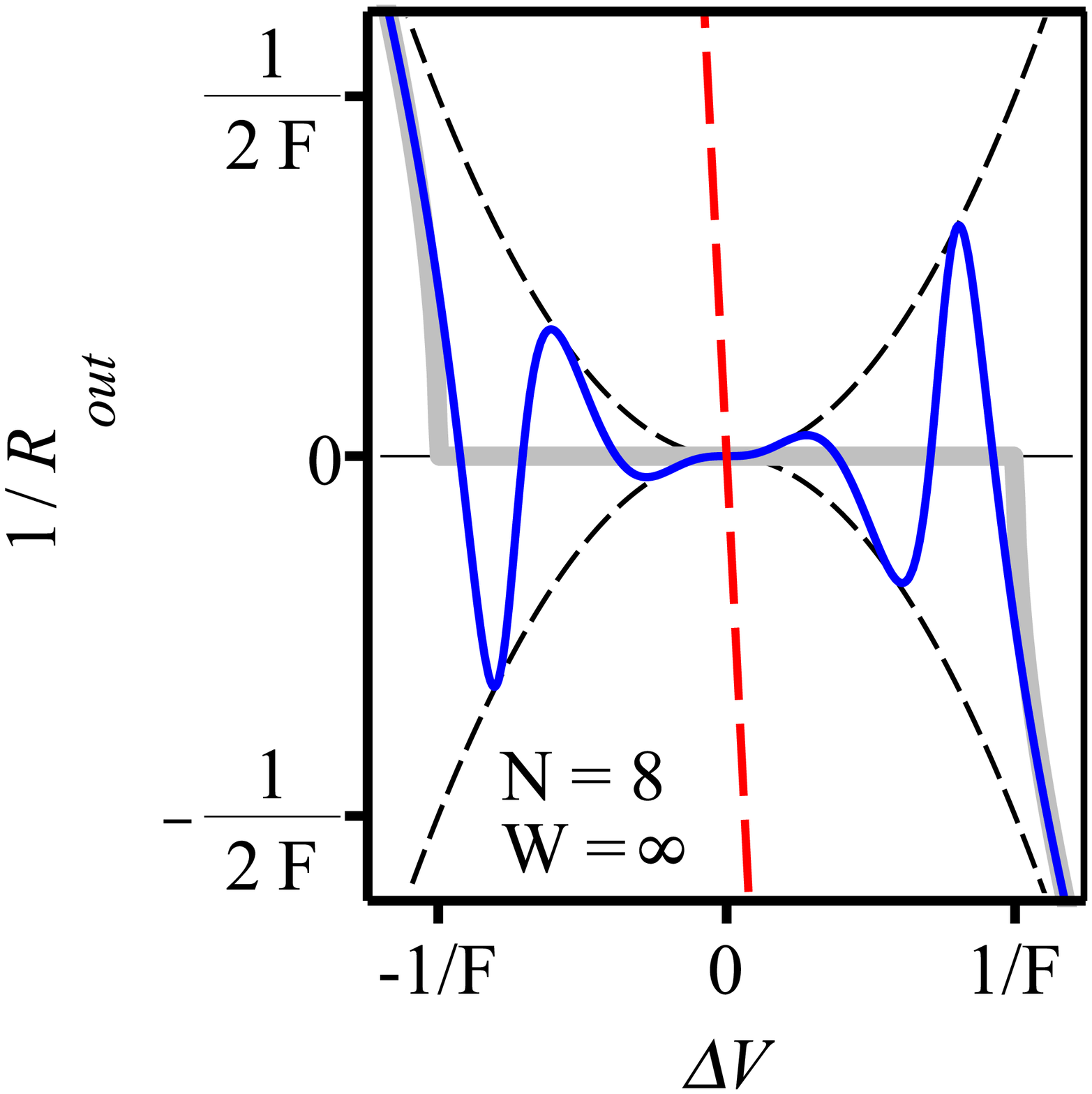} 
    \caption{} \label{Curvature3_40}
  \end{subfigure}
  \caption{Similar to Fig.~\ref{Fig3} but for an 8-pass amplifier as given in Figs.~\ref{fig:multy-pass-2f-only} and~\ref{fig:multi-passes}. 
The black dashed lines represent the maximally allowed deviations for vanishing aperture effects given by Eqs.~(\ref{eq:bound_w}) and    (\ref{eq:bound_R}).  }
\label{Fig7}
\end{figure}

\begin{figure}[t!]
\centering
  \begin{subfigure}[t]{.45 \linewidth}
    \centering
    \includegraphics[width= \linewidth]{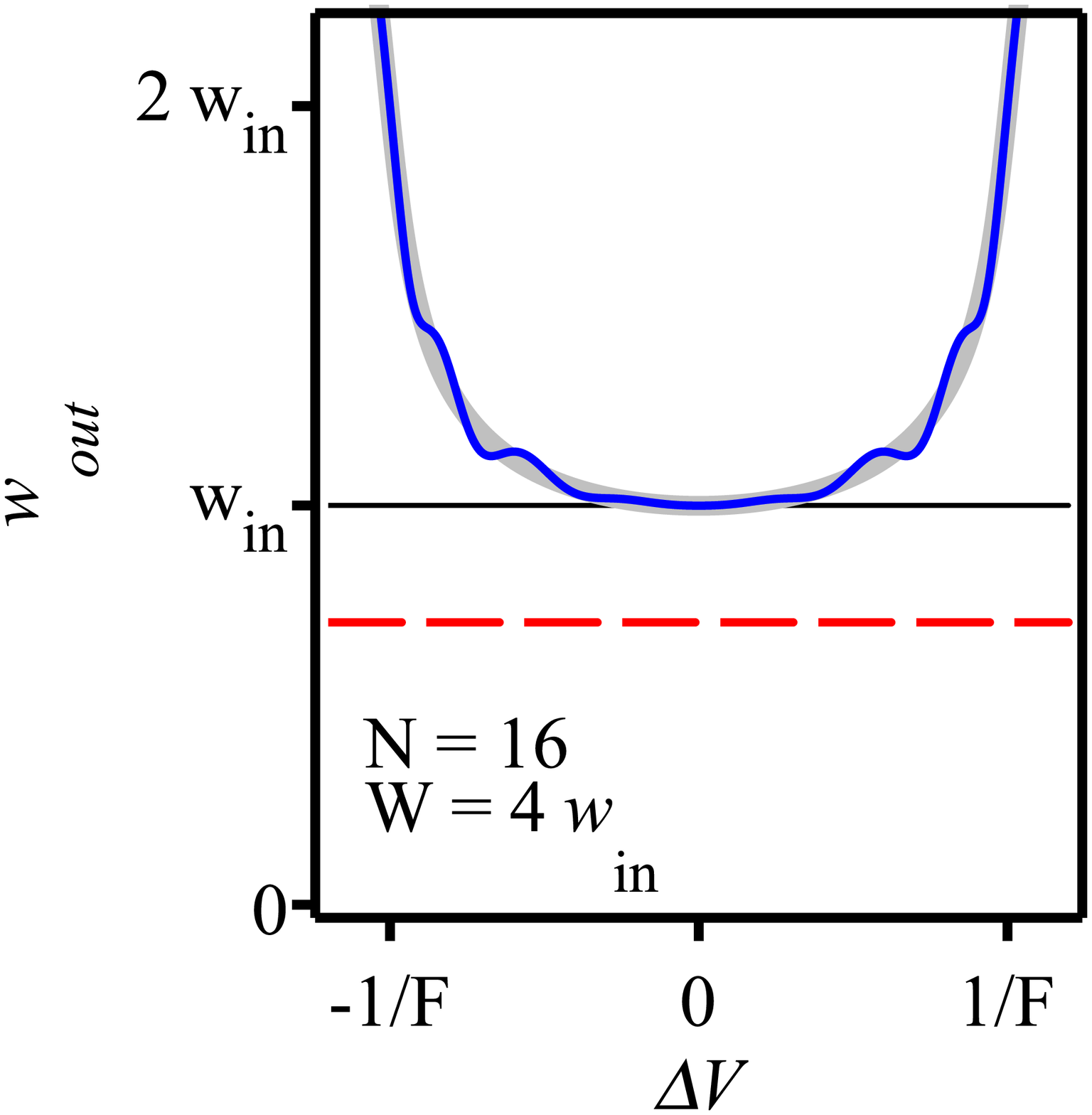} 
    \caption{} \label{Waist4}
  \end{subfigure}
  \hfill
  \begin{subfigure}[t]{.45 \linewidth}
    \centering
    \includegraphics[width= \linewidth]{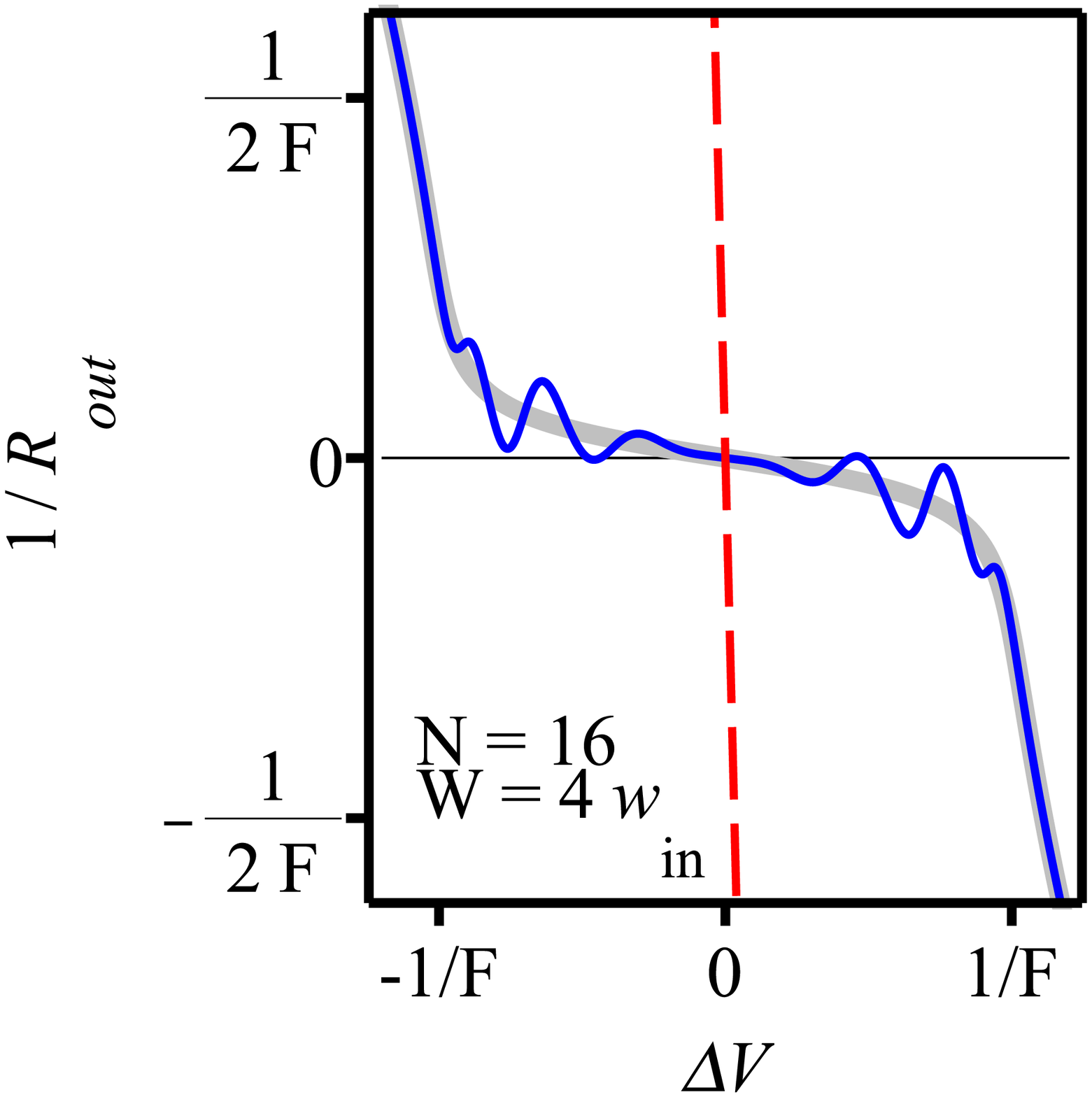} 
    \caption{} \label{Curvature4}
  \end{subfigure}
    \begin{subfigure}[t]{.45 \linewidth}
    \centering
    \includegraphics[width= \linewidth]{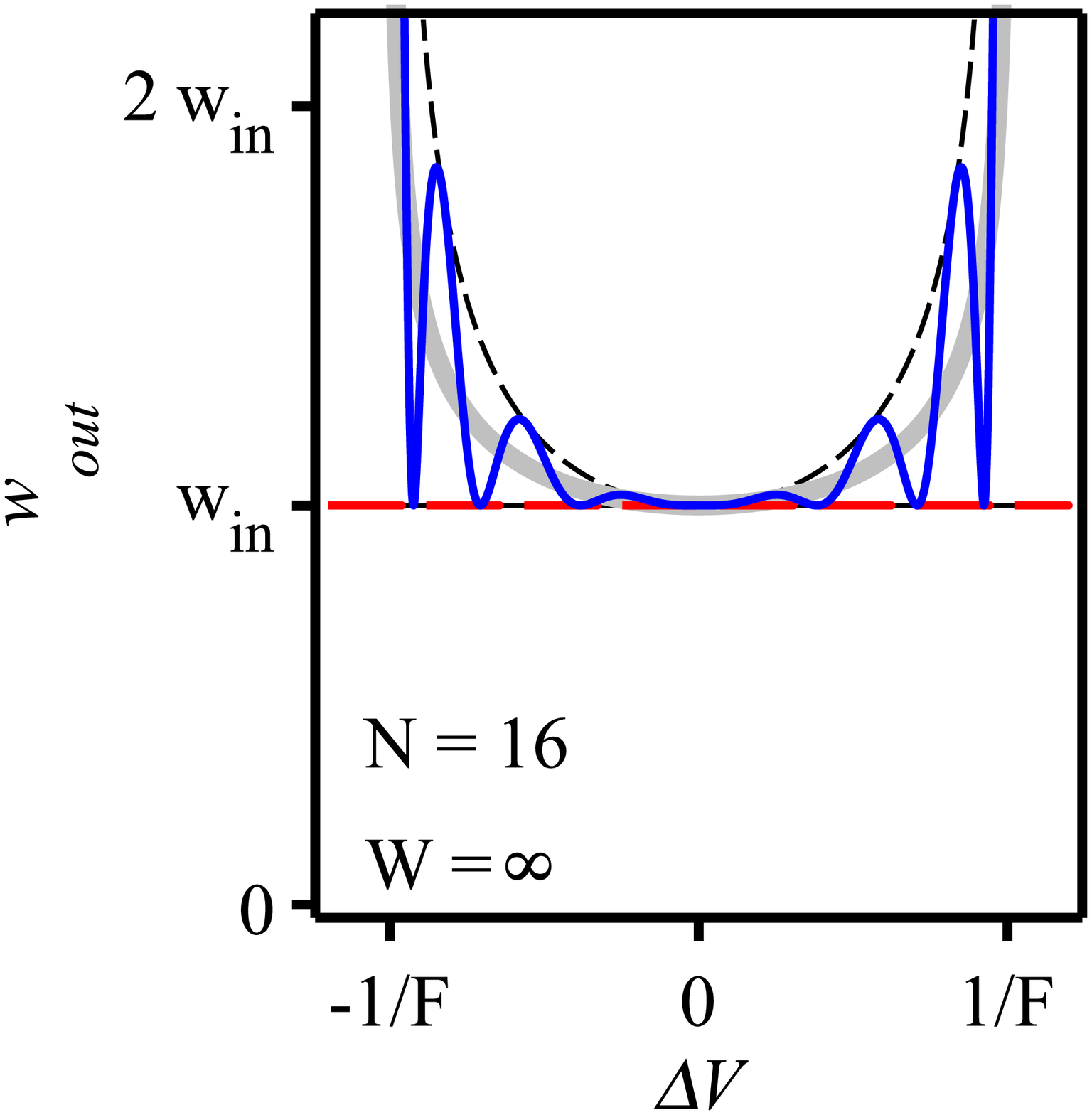} 
    \caption{} \label{Waist4_40}
  \end{subfigure}
  \hfill
  \begin{subfigure}[t]{.45 \linewidth}
    \centering
    \includegraphics[width= \linewidth]{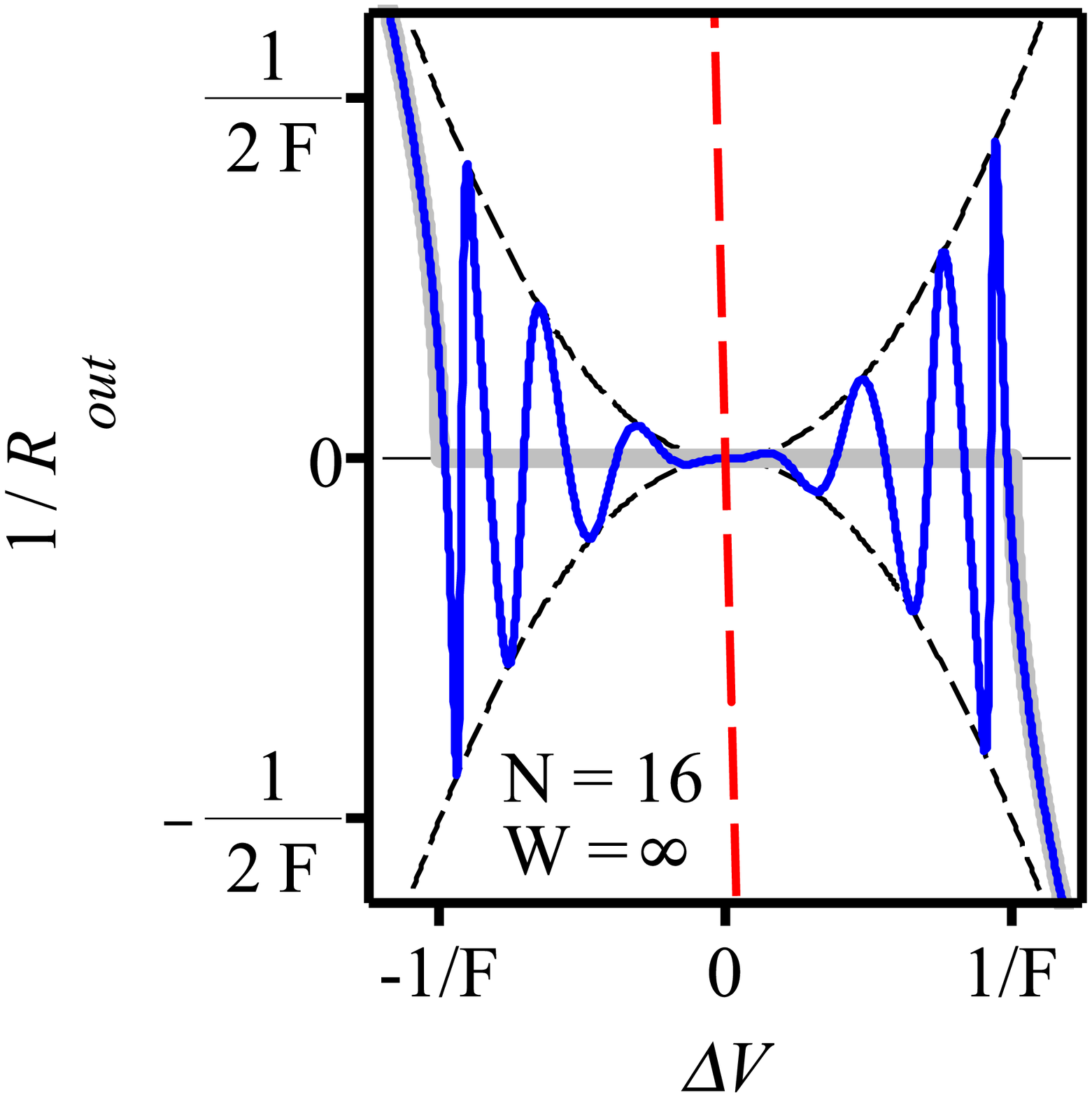} 
    \caption{} \label{Curvature4_40}
  \end{subfigure}
  \caption{Similar to Fig.~\ref{Fig7} but for a 16-pass amplifier.  }
\label{Fig7_32passes}
\end{figure}

 
An example of an 8-pass amplifier modeled according to this scheme is shown in Fig.~\ref{fig:multy-pass-2f-only}.
Size and phase front curvature of the output beam for an 8-pass and a 16-pass amplifier as a function of $\Delta V$ are shown in Figs.~\ref{Fig7} and ~\ref{Fig7_32passes}, respectively.
The accumulation of lens and aperture effects in the 4f-based amplifiers becomes striking when considering the red dashed curves in these figures. In such multi-pass designs, the size of the output beam $w_\mathrm{out}$ becomes significantly smaller than the input beam size $w_\mathrm{in}$, while the phase front curvature $1/R_\mathrm{out}$ of the output beam becomes increasingly sensitive to variations of $\Delta V$ (large slope of the red curves).

Contrarily, in the Fourier-based amplifiers the output beam characteristics show an oscillatory behavior around the design value for variations of $\Delta V$. 
For vanishing aperture effects this oscillation is bound (see black dashed lines in Figs.~\ref{Fig7} and \ref{Fig7_32passes}) by the following relations:
\begin{eqnarray}
 w_\mathrm{in} < \, \, & w_\mathrm{out, \, N}(\Delta V) \,\, &< \,\, w_\mathrm{stable}^2(\Delta V)  /w_\mathrm{in} \ , \label{eq:bound_w} \\
-\frac{F}{2}\Delta V^2 < \, \, &R^{-1}_\mathrm{out, \, N}(\Delta V)  \,\, &< \,\, \frac{F}{2}\Delta V^2 \ ,
\label{eq:bound_R}
\end{eqnarray}
where $w_\mathrm{stable}(\Delta V)$ and $R^{-1}_\mathrm{stable}(\Delta V) $ are the size and phase front radius obtained from Eq.~(\ref{eq:condition-2a}) represented by the gray curves in Figs.~\ref{Fig7} and ~\ref{Fig7_32passes}.
A non-vanishing aperture in the active medium leads to a damping of this oscillatory behavior so that the output beam characteristics approaches the design value corresponding to the complex beam parameter $q_\mathrm{stable}$ given in Eq.~(\ref{eq:condition-2a}).
In this case the gray curves represent the output beam characteristics for an infinite number of passes.
A similar damping related with aperture effects has been discussed in detail also in the context of multi-pass oscillators \cite{Schuhmann2017}.

The Taylor expansion of the complex beam parameter after a propagation through a Fourier-based 8-pass amplifier is
\begin{eqnarray}
  q_\mathrm{out, \, 8} &\hspace{-3mm}=&\hspace{-3mm}  iF +4\,F^4\Delta \tilde{V}^3
                      +16\,iF^5 \Delta \tilde{V}^{4}
                      -40\,F^6  \Delta \tilde{V}^{5}
                      + \cdots \ ,
\label{eq.q_8}
\end{eqnarray}
to be compared with the beam parameter after a 32-pass propagation
\begin{eqnarray}
q_\mathrm{out, \, 32} &\hspace{-3mm}=&\hspace{-3mm} iF+16\,{F}^{4}{\Delta  \tilde{V}}^{3}+256\,i{F}^{5}{\Delta  \tilde{V}}^{4}-2720\,{F}^{6}{\Delta  \tilde{V}}^{5} \cdots \ .
\end{eqnarray}

From these equations, the corresponding output beam sizes and phase front curvatures can be deduced for variations of $\Delta V$. Assuming $W=\infty$ the results are
\begin{eqnarray}
  \frac{w_\mathrm{out, \, 8}}{w_\mathrm{in}} &\hspace{-3mm} = &\hspace{-3mm} 1+ 8{F}^{4}{\Delta V}^{4}-32{F}^{6}{\Delta V}^{6} + \cdots  ,\\
  \frac{w_\mathrm{out, \, 32}}{w_\mathrm{in}}  &\hspace{-3mm} = &\hspace{-3mm}  1+128{F}^{4}{\Delta V}^{4}-10752{F}^{6}{\Delta V}^{6}+\cdots , \\    
R^{-1}_\mathrm{out, \, 8}&\hspace{-3mm}=& \hspace{-3mm} 4\,{F}^{2}{\Delta V}^{3}-40\,{F}^{4}{\Delta V}^{5}+32\,{F}^{6}{\Delta V}^{7}+ \cdots \ , \\
R^{-1}_\mathrm{out, \, 32} &\hspace{-3mm}=&\hspace{-3mm} 16F^2 \Delta V^3
-2720 F^4 \Delta V^5 + 132992F^6 \Delta V^7\cdots  \ .
\end{eqnarray}
Similar to the Fourier-based 4-pass amplifiers, the output beam waist and divergence of the 8-, 16-, 32-pass amplifiers depend in lowest order on $\Delta V^4$ and $\Delta V^3$, respectively.
The coefficients in front of  $\Delta V^4$ and $\Delta V^3$ increases with the number of passes but still the maximal deviation from the design value (gray curves) is bound by the relations of Eqs.~(\ref{eq:bound_w}) and (\ref{eq:bound_R}).

Alike previous sections we have investigated  the dependence of the output beam size on the aperture size $W$.
Assuming for simplicity that we are in the center of the ``stability region'' ($V=\Delta V=0$) we find that
\begin{eqnarray}
  \frac{w_\mathrm{out, \, 8}}{w_\mathrm{in}}  &\hspace{-3mm} = &\hspace{-3mm}  1-\frac{2w_\mathrm{in}^6}{W^6}  
  +\frac{8\,w_\mathrm{in}^8}{W^8} +  \cdots  \ ,\\
  \frac{w_\mathrm{out, \, 32}}{w_\mathrm{in}}  &\hspace{-3mm} = &\hspace{-3mm}  1-\frac{8w_\mathrm{in}^6}{W^6} +\frac{128 w_\mathrm{in}^8}{W^8}+\cdots \ ,
\label{eq.w_32}
\end{eqnarray}
and that the phase front curvature of the output beam does not depend on the aperture size. 
Unlike the behavior for a 4f-based amplifier, the beam waist is converging towards the values of Eq.~(\ref{eq:condition-2a}).
The converging behavior that is apparent in the top panels of Figs~(\ref{Fig7}) and (\ref{Fig7_32passes}) is not obviously manifest from the Taylor expansions of Eqs.~(\ref{eq.q_8}) to (\ref{eq.w_32}).

\section{Practical realization}
\label{sec:implementation}

The Fourier-based multi-pass amplifiers sketched in the previous sections show excellent output beam stability for variations of $\tilde V$ but their practical implementation as such, leads to long propagation lengths.
In fact, the large mode size at the active medium needed for high-power and high-energy operation requires the use of large focal lengths $F$:
\begin{equation}
  F = \frac{\pi \, w_\mathrm{in}^2}{\lambda} \ .
  \label{eq:F_vs_waist}
\end{equation}
This result was obtained by solving Eq.~(\ref{eq:condition-2}) for $F$.
For example, a beam size at the active medium of 1~mm requires $F = 3$~m for $\lambda = 1030$~nm, so that each Fourier transform is 6~m long.

At the same time, it is challenging to realize an active medium with vanishing dioptric power $V=V_\mathrm{unpumped}+V_\mathrm{thermal}^\mathrm{avg}=0$ at the desired operational conditions as assumed in previous sections.

The realization of the Fourier transform propagation using two pairs of Galilean telescopes solves both these issues: it shortens the physical propagation length and allows for the use of active media with non-vanishing dioptric power.
However, the focal lengths and the positions of the mirrors forming the telescope pairs must be chosen  so that the propagations from active medium to active medium correspond approximatively to a Fourier transform, i.e. their ABCD-matrices must have the form $A=D=0$ and $B=1/C=-F$.

\begin{figure}[t!]
\centering
  \begin{subfigure}[t]{.45 \linewidth}
    \centering
    \includegraphics[width= \linewidth]{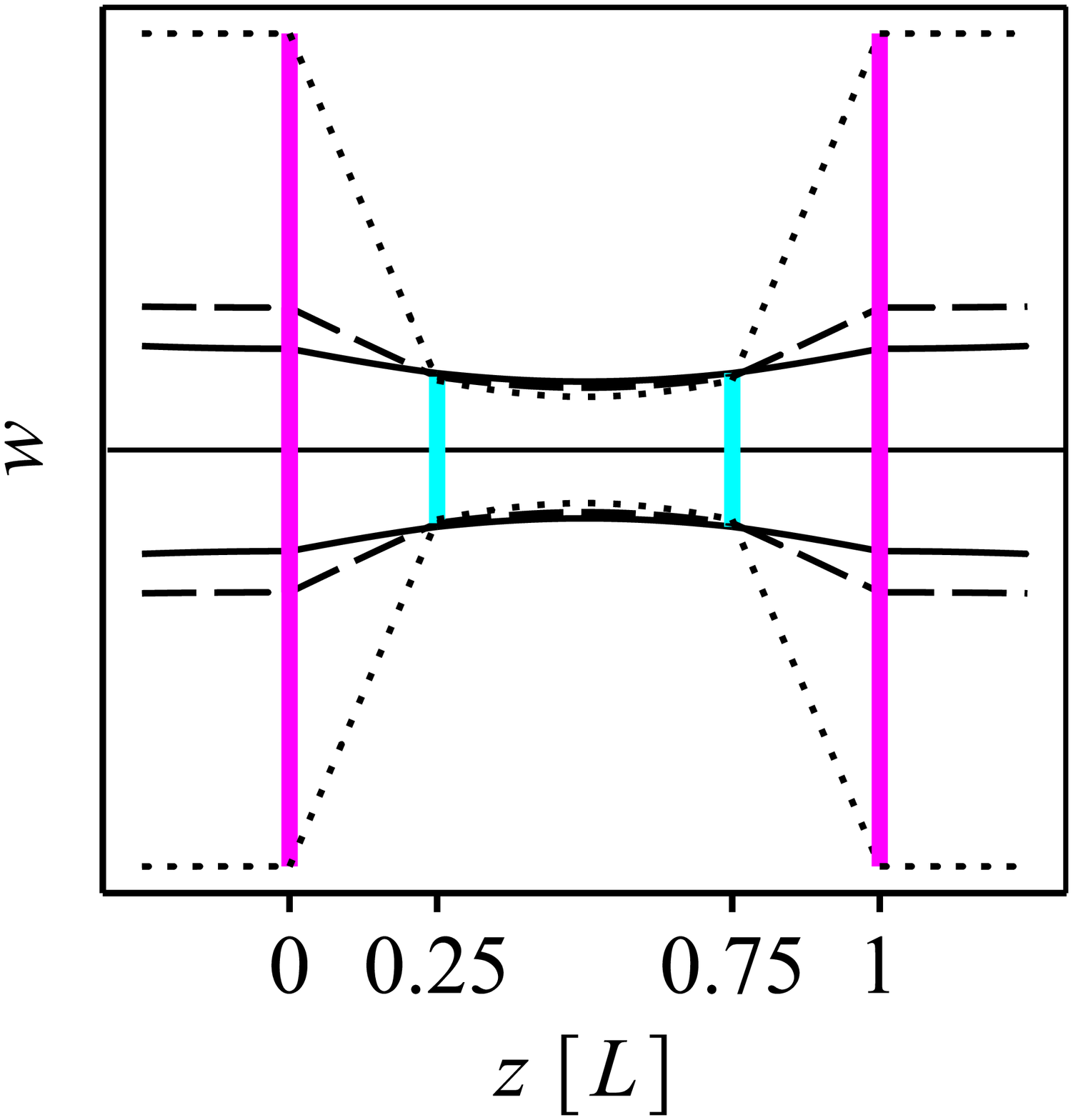} 
    \caption{} \label{Telescope}
  \end{subfigure}
  \hfill
  \begin{subfigure}[t]{.47 \linewidth}
    \centering
    \includegraphics[width= \linewidth]{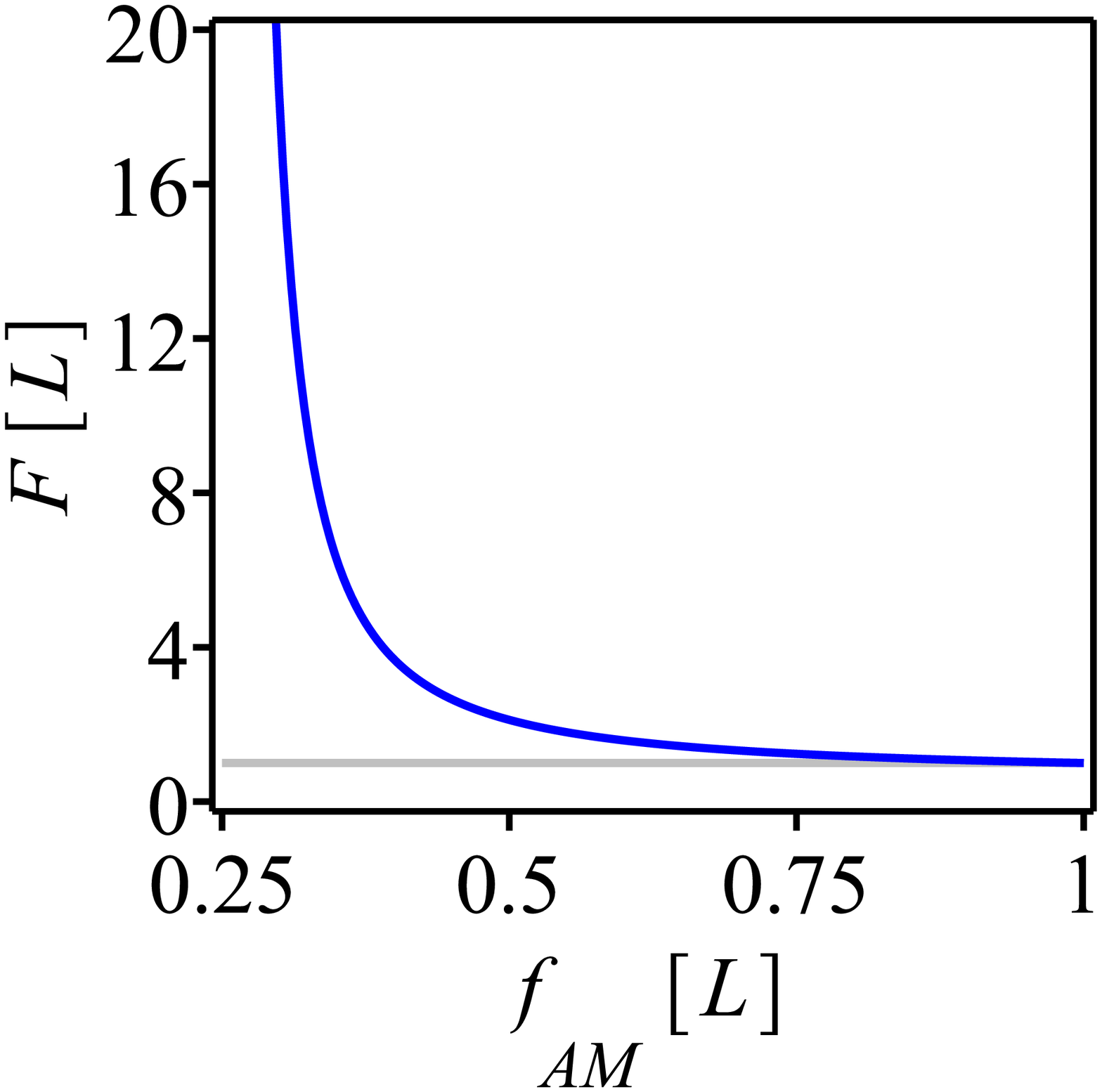} 
    \caption{} \label{Stretching}
  \end{subfigure}
  \caption{(a) Scheme of the optical Fourier transform from active medium to active medium obtained using two Galilean telescopes. The active medium itself (vertical lines at $z=0$ and $z=1$) acts as focusing element of the telescopes with focal length $f_{AM}$. The defocusing elements (cyan vertical lines) of the telescopes have focal length $f'$ given by Eq.~(\ref{eq:f'}) and are placed at a distance $0.25L$ from the active medium, where $L$ is the distance between the two passes in the active medium. The beam size ($w$) evolution in the Fourier    transform segment is given for 3 different values of the active medium focal length: $f_{AM}= L$ (solid), $f_{AM} = 0.5 L$ (dashed) and $f_{AM} = 0.3 L$ (dotted). (b) Effective focal length $F$ of the Fourier transform  as a function of the active medium focal length $f_{AM}$ expressed in units of $L$. }
\label{Fig8}
\end{figure}

In the thin-disk lasers sector it is customary to use thin disks (active media) with a focusing focal length ($f_{AM}$).
Therefore, as shown in Fig.~\ref{Telescope}, the active medium itself can be used as curved mirror to realize the Galilean telescopes.
To fully define the layout of the Fourier transform, we need to further assume the position of the defocusing optical elements (lens or mirrors) forming the telescope pairs.
In the layout shown in Fig.~\ref{Telescope}, it was assumed that these defocusing elements were placed at a distance $0.25L$ after and before the active media, where $L$ is the distance from active medium to active medium.
Under these assumptions, we find that the telescopic pair act as a Fourier transform when the focal length $f'$ of these defocusing elements fulfill the relation
\begin{equation}
f'=\frac{1}{2}\, \frac { \left( -4\,f_{AM}+L \right) L}{3\,L-8\,f_{AM}+\sqrt {{L}^
    {2}-8\,Lf_{AM}+32\,{f_{AM}}^{2}}}  \ .
\label{eq:f'}
\end{equation}
The resulting equivalent value of $F$ takes then the form
\begin{equation}
F=f_{AM}\frac{\sqrt{32\ {\left(f_{AM}/L\right)}^2-8\ {f_{AM}}/{L}+1}+4{f_{AM}}/{L}}{{\left(4{f_{AM}}/{L-1}\right)}^2}
\ .
\label{eq:F-effective}
\end{equation}
As can be seen from Fig.~\ref{Stretching}, the ratio $F/L$, which describes the physical shortening factor of the Fourier transform length due to the implementation of the Galilean telescopes, rapidly increases with decreasing $f_{AM}/L$ ratio.
Thus, a small value of $f_{AM}/L$ via Eqs.~(\ref{eq:F_vs_waist}) and (\ref{eq:F-effective}) leads to a large beam size at the active medium even for a short physical propagation length $L$.

In the practical implementation however, there is a limit in the minimal value of the focal length $f_\mathrm{AM}$ given by the optical damage at the defocusing elements and the laser-induced breakdown in air.
Indeed the beam sizes at the defocusing elements and in the focus of the telescopic system decrease with decreasing $f_{AM}$.
\begin{figure}[t!]
\centering
  \begin{subfigure}[t]{.9 \linewidth}
    \centering
    \includegraphics[width= \linewidth]{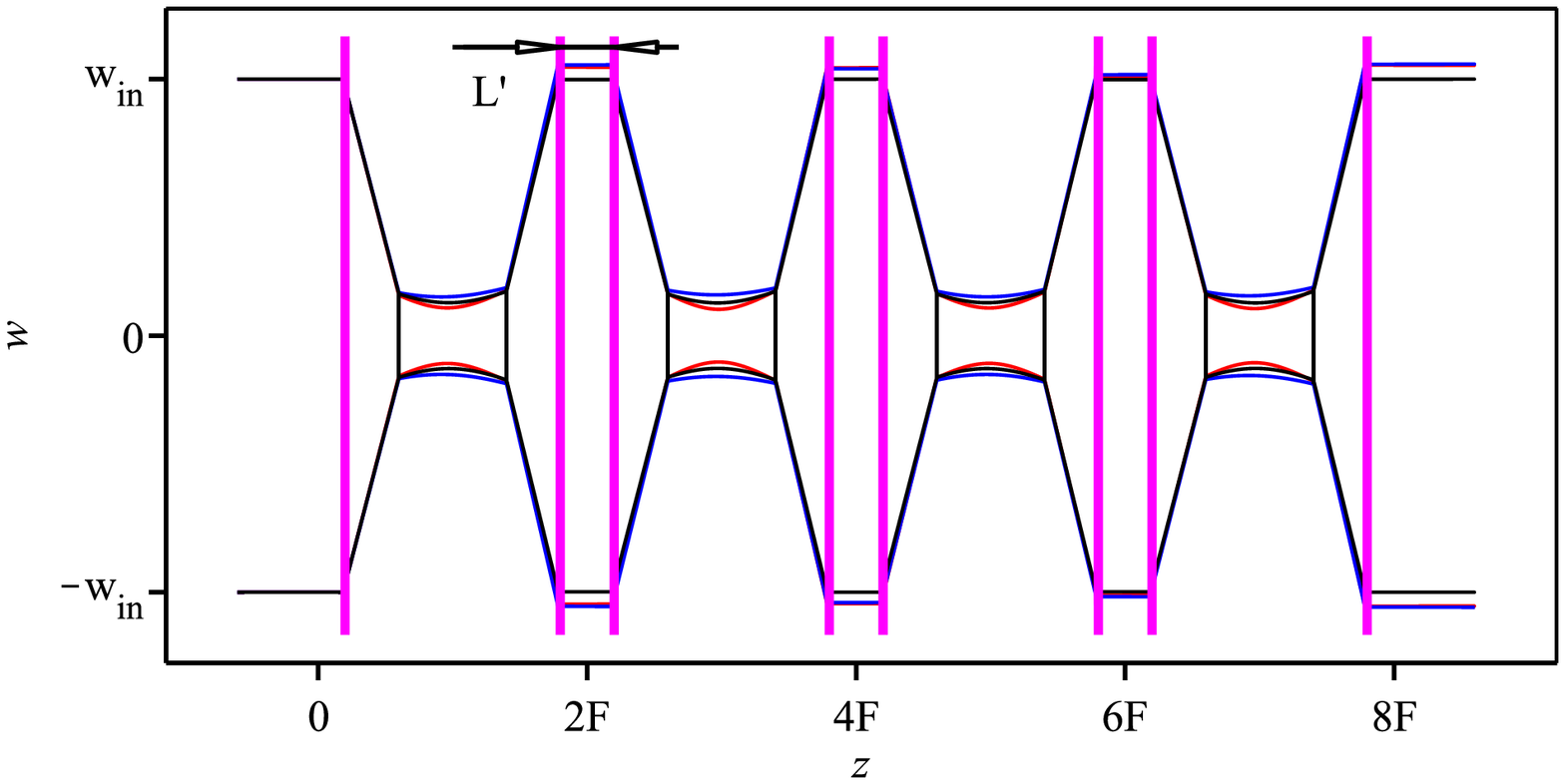} 
    \caption{} \label{multi-pass-1}
  \end{subfigure}
  \hfill
  \begin{subfigure}[t]{.9 \linewidth}
    \centering
    \includegraphics[width= \linewidth]{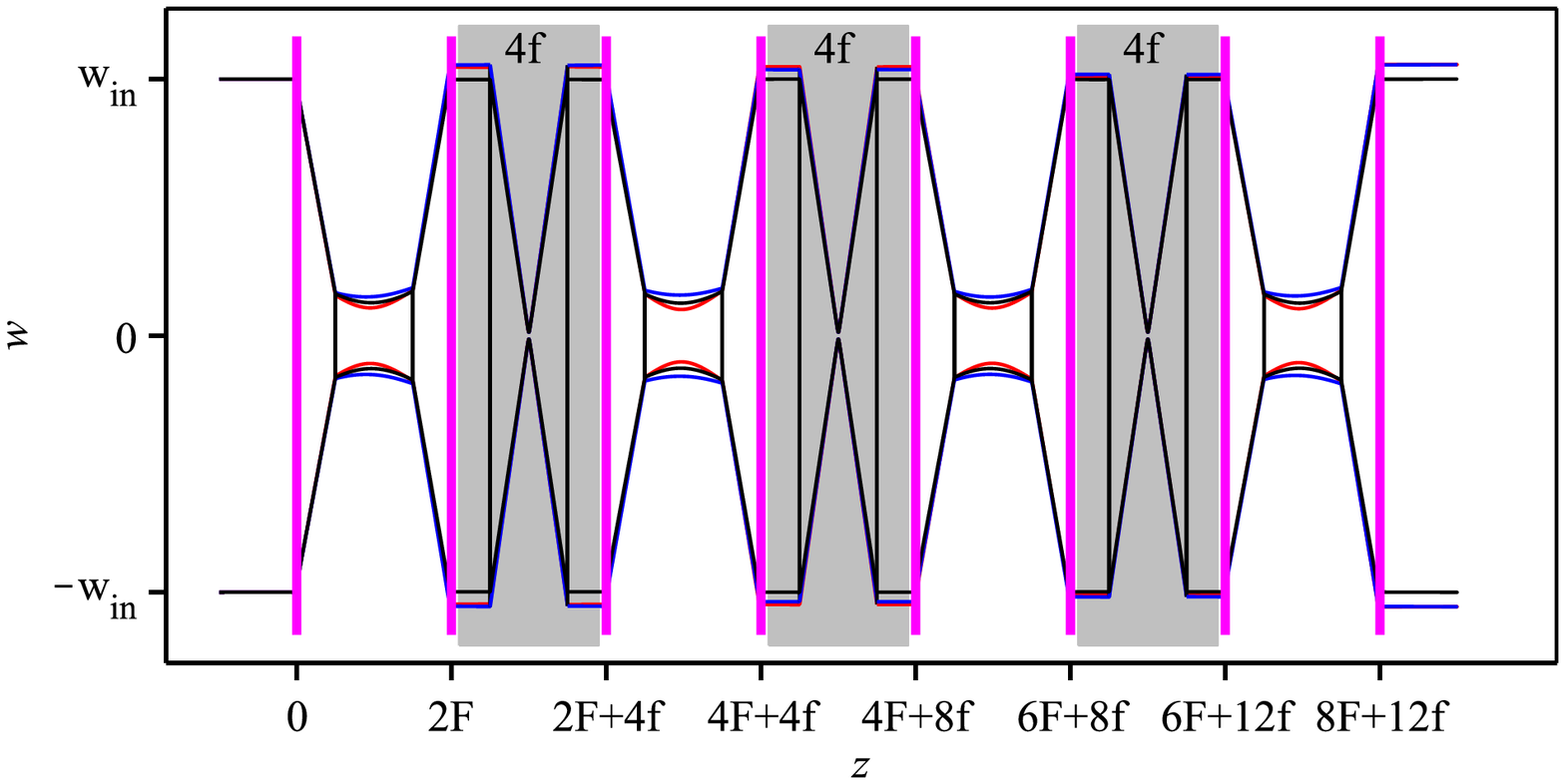} 
    \caption{} \label{multi-pass-2}
  \end{subfigure}
\caption{(a) Scheme of an 8-pass Fourier-based amplifier obtained by concatenating two 4-pass amplifiers of Fig.~\ref{Fig4}.
 Each Fourier transform has been shortened using two Galilean telescopes. 
 The active medium is also part of the Galilean telescope.
 Between two Fourier transforms, there is a short propagation of length $L'$. (b) Similar to (a) but in this case the 8-pass  amplifier is based on Fig.~\ref{Fig5}. 
 Between each Fourier transforms, there is a 4f-imaging (indicated by the gray shaded area). 
 Same notation as in Fig.~\ref{fig:4f-segment} is used.}
\label{fig:multi-passes}
\end{figure} 

We conclude by sketching the optical layout and the corresponding beam size evolution for two 8-pass Fourier-based amplifiers realized with the above-described Galilean telescopes.
In these designs the active medium that has a focusing dioptric power $V$ is also utilized as focusing element of the Galilean telescope.
In Fig.\ref{multi-pass-1}, the propagation from active medium to active medium follows the sequence
\begin{equation*}
AM-Fourier-AM-L'-AM-Fourier-AM-L'\cdots \ ,
\end{equation*}
where $L'$ is a short propagation distance compared to the Rayleigh length of the beam at the active medium, while in Fig.~\ref{multi-pass-2} the propagation from active medium to active medium follows the sequence
\begin{equation*}
AM-Fourier-AM-4f-AM-Fourier-AM-4f \cdots \ ,
\end{equation*}
i.e., an alternation between Fourier- and 4f-propagations.
The optical properties of the output beam for both these sequences corresponding to the two amplifier layout of Fig.~\ref{fig:multi-passes} are shown in Fig.~\ref{Fig7}.

The shortening of the propagation due to the insertion of telescopes can be better appreciated with a numerical example.
A large focal length of $F=90$~m (see Eq.~\ref{eq:F_vs_waist}) is needed to realize a multi-pass amplifier based on the scheme of Fig.~\ref{fig:multy-pass-2f-only}  (without telescopes) and having a beam size $w_\mathrm{in}= 5.4$~mm and a wavelength of $\lambda = 1030$~nm.  
Assuming the short propagation length is $L'= 1$~m the total length of the 8-pass amplifier would be $L_\mathrm{total} = 720$~m.
Using the layout of Fig.~\ref{multi-pass-1}, the same beam size at the active medium can be obtained for an active medium with focal length $f_\mathrm{AM}=1$~m, and a defocusing element with focal length $f'=-150$~mm placed at a distance of 861~mm. 
In this way the distance AM-Fourier-AM is reduced to $3.445$~m (instead of 180~m). 
Assuming that $L'= 1$~m the total length of the 8-pass amplifier is reduced to $L_\mathrm{total} = 16.8$~m (instead of 720~m).
Because this telescope reduces the beam size in the mid-plane of the Fourier transform to $w=0.57$~mm, optical damage issues have to be considered.
Yet, this beam size is significantly larger than the size of $w=0.06$~mm in the focal plane of a 4f-imaging with  $f=1000$~mm.

\section{Conclusion}

Numerous studies have investigated the influence of the active medium thermal lens on resonator (oscillator) eigenmodes~\cite{Kogelnik:65,Magni1986}.
Oscillator designs have been developed that mitigate this influence reducing the dependence of the oscillator performance and output beam characteristics to variations of the thermal lens of the active medium.
However, similar studies have not (or insufficiently) been undertaken for multi-pass  amplifiers.
This might be associated to the fact that the 4f-relay imaging, that is commonly used when realizing multi-pass amplifiers, offers two obvious advantages.
The first is that its imaging properties reproduce the same transverse intensity profile (size) from pass to pass independently on the size and shape of the beam and  independently of the active medium dioptric power.
The second, is that several passes can be implemented with only few optical elements.
Less obvious and known are the disadvantages. 
The first is that the beam size is reproduced only for a vanishing aperture effect at the active medium. 
Secondly, there is a pile up of phase front distortions at each pass on the active medium that strongly affects the output beam divergence and the beam quality.
Imaging based amplifiers are thus well suited for low-power lasers where the thermal lens is small, but become less apt for high-power lasers with good output beam quality.

Here we have presented a multi-pass amplifier architecture based on a succession of optical Fourier propagations.
Contrarily to the 4f-imaging the layout of this Fourier transform propagation needs to be designed to match the desired beam size at the active medium.
Moreover, the implementation of this architecture requires one additional mirror for every pass, making it more complex compared to the 4f-based layouts.
However, the implementation of these minor additional complexities is rewarded with the compensation of the thermal lens effects occurring in this architecture which significantly decreases the sensitivity of size and
divergence of the output beam to variations of the dioptric power and aperture effects of the active medium.
Therefore, this architecture overcomes present energy- and power-scaling limitations related with thermal lens effect in the active medium.
Furthermore, the mode filtering that naturally occurs from the interplay between the Fourier transform propagations and the soft aperture of the active medium provides excellent beam quality with high efficiency.

We found that the output beam size and phase front radius of a Fourier-based multi-pass system are less sensitive to variations of the active medium thermal lens, than a single-pass amplifier.
Hence, from the point of view of the stability to thermal lens effects, it is advantageous to realize amplifiers as multi-pass amplifiers where the propagating beam is redirected several times to the same active medium.
Various active media can be utilized provided they have identical or similar dioptric power.
The multi-pass architecture presented here is thus particularly suited for thin-disk lasers where the gain per pass is small and multiple passes are required.
An example of such a multi-pass amplifier realized according to this architecture is presented in~\cite{5165103, Schuhmann2015} where the disk is acting as active medium, as soft aperture and as focusing mirror of the Galilean telescopes used to realize the Fourier transform.

We want to point out that the here presented architecture can be used in other contexts as well.
For example, a recent development in Faraday insulators reduced the thermal lens effect resulting in TEM00 up to 400~W average power \cite{LATJ:LATJ201600017, Pavos_Ultra}.
However, commercial available laser systems already provide output powers exceeding 10~kW \cite{High_Power_CW_Fiber_Lasers}.
In the laser system at advanced LIGO the focusing thermal lens effect of TGG (terbium gallium granate) used in the Faraday rotator was compensated using a plate of DKDP (deuterated potassium dihydrogen phosphate) exhibiting a defocusing thermal lens~\cite{Zelenogorsky}.
Following our design (as presented in Fig.~\ref{Fourier}) the effective thermal lens of the Faraday rotator can be almost canceled (see Fig.~\ref{Fig3}) without additional transmissive elements simply by splitting the rotator in two halves and by implementing a Fourier transform between them.

\section*{Funding Information}

We acknowledge the support from the Swiss National Science Foundation Project SNF 200021\_165854, the European Research Council ERC CoG. \#725039 and ERC StG. \# 279765. 
The work was supported by ETH Research Grant ETH-35 14-1.
The study has been also supported by the ETH Femtosecond and Attosecond Science and Technology (ETH-FAST) initiative as part of the NCCR MUST program.

\bibliographystyle{unsrt}
\bibliography{sample}

\end{document}